\documentclass[acp, manuscript]{copernicus_v}

\begin{document}

\title{Structure, variability, and origin of the low-latitude nightglow
  continuum between 300 and 1,800\,\unit{nm}: Evidence for \chem{HO_2}
  emission in the near-infrared}


\Author[1,2]{Stefan}{Noll}
\Author[3]{John M.~C.}{Plane}
\Author[3,4]{Wuhu}{Feng}
\Author[5]{Konstantinos S.}{Kalogerakis}
\Author[6]{Wolfgang}{Kausch}
\Author[2]{Carsten}{Schmidt}
\Author[2,1]{Michael}{Bittner}
\Author[6,7]{Stefan}{Kimeswenger}

\affil[1]{Institut für Physik, Universit\"at Augsburg, Augsburg, Germany}
\affil[2]{Deutsches Fernerkundungsdatenzentrum, Deutsches Zentrum f\"ur Luft-
  und Raumfahrt, Oberpfaffenhofen, Germany}
\affil[3]{School of Chemistry, University of Leeds, Leeds, UK}
\affil[4]{National Centre for Atmospheric Science, University of Leeds, Leeds,
  UK}
\affil[5]{Center for Geospace Studies, SRI International, Menlo Park, CA, USA}
\affil[6]{Institut f\"ur Astro- und Teilchenphysik, Universit\"at Innsbruck,
  Innsbruck, Austria}
\affil[7]{Instituto de Astronom\'ia, Universidad Cat\'olica del Norte,
  Antofagasta, Chile}




\correspondence{S. Noll (stefan.noll@dlr.de)}

\runningtitle{The nightglow continuum}

\runningauthor{S. Noll et al.}

\received{}
\pubdiscuss{} 
\revised{}
\accepted{}
\published{}


\firstpage{1}

\maketitle

\begin{abstract}
  The Earth's mesopause region between about 75 and 105\,\unit{km} is
  characterised by chemiluminescent emission from various lines of different
  molecules and atoms. This emission was and is important for the study of
  the chemistry and dynamics in this altitude region at nighttime. However,
  our understanding is still very limited with respect to molecular emissions
  with low intensities and high line densities that are challenging to
  resolve. Based on 10 years of data from the astronomical \mbox{X-shooter}
  echelle spectrograph at Cerro Paranal in Chile, we have characterised in
  detail this nightglow (pseudo-)continuum in the wavelength range from 300 to
  1,800\,\unit{nm}. We studied the spectral features, derived continuum
  components with similar variability, calculated climatologies, studied the
  response to solar activity, and even estimated the effective emission
  heights. The results indicate that the nightglow continuum at Cerro Paranal
  essentially consists of only two components, which exhibit very different
  properties. The main structures of these components peak at 595 and
  1,510\,\unit{nm}. While the former was previously identified as the main
  peak of the \chem{FeO} `orange arc' bands, the latter is a new discovery.
  Laboratory data and theory indicate that this feature and other structures
  between about 800 and at least 1,800\,\unit{nm} are caused by emission from
  the low-lying A$^{\prime\prime}$ and A$^{\prime}$ states of \chem{HO_2}. In
  order to test this assumption, we performed runs with the Whole Atmosphere
  Community Climate Model (WACCM) with modified chemistry and found that the
  total intensity, layer profile, and variability indeed support this
  interpretation, where the excited \chem{HO_2} radicals are mostly produced
  from the termolecular recombination of \chem{H} and \chem{O_2}. The WACCM
  results for the continuum component that dominates at visual wavelengths
  show good agreement for \chem{FeO} from the reaction of \chem{Fe} and
  \chem{O_3}. However, the simulated total emission appears to be too low,
  which would require additional mechanisms where the variability is dominated
  by \chem{O_3}. A possible (but nevertheless insufficient) process could be
  the production of excited \chem{OFeOH} by the reaction of \chem{FeOH} and
  \chem{O_3}.
\end{abstract} 


\introduction[Introduction]  
\label{sec:intro}

At wavelengths shorter than about 1,800\,\unit{nm}, the Earth's atmospheric
radiation at nighttime is essentially caused by non-thermal chemiluminescence,
i.e. photon emission by excited atomic and molecular states that are populated
as a result of chemical reactions. Most of this nightglow emission originates
at altitudes between 75 and 105\,\unit{km} in the mesopause region. The most
prominent emitting species are the hydroxyl radical (\chem{OH}) and molecular
oxygen (\chem{O_2}), which cause various ro-vibrational bands of emission
lines from the near-ultraviolet (near-UV) to the near-infrared (near-IR)
\citep{rousselot00,cosby06,noll12}. Especially strong emission is found above
1,400\,\unit{nm}, where \chem{OH} bands of the electronic ground level with a
vibrational level change $\Delta v$ of 2, e.g. \mbox{\chem{OH}(3-1)}, are
located. Bands with higher $\Delta v$ that can be found at shorter wavelengths
are significantly weaker. Strong emission is also related to
\mbox{\chem{O_2}(b-X)(0-0)} near 762\,\unit{nm} and
\mbox{\chem{O_2}(a-X)(0-0)} near 1,270\,\unit{nm}. However, both bands suffer
from strong self-absorption in the lower atmosphere, which makes it
particularly challenging to observe any emission of the former band from the
ground. Intrinsically weaker but not self-absorbed \chem{O_2} bands are
\mbox{(b-X)(0-1)} near 865\,\unit{nm} and \mbox{\chem{O_2}(a-X)(0-1)} near
1,580\,\unit{nm}. Moreover, there are many weak \chem{O_2} bands at near-UV
and blue wavelengths \citep{slanger03,cosby06}. In addition, especially the
visual range shows atomic emission lines. Prominent examples are the atomic
oxygen (\chem{O}) lines at 558, 630, and 636\,\unit{nm} and the sodium
(\chem{Na}) doublet at 589\,\unit{nm} \citep[e.g.,][]{cosby06,noll12}.

Apart from individual emission lines, which have a typical width of a few
picometres, the nightglow also includes an underlying continuum component.
It could consist of line emissions if there were such a high line density
that even spectroscopic instruments with high resolving power were not able to
distinguish individual lines, i.e. it would be a pseudo-continuum. In any
case, the observation of such a (pseudo-)continuum is more challenging than
the study of well-resolved emission lines. The applied instrument needs to
have a sufficiently high resolving power to clearly separate the continuum
from the well-known emission bands and lines. As the continuum can be quite
faint compared to the strong lines, even the wings of the line-spread function
and possible straylight inside the instrument can be an issue, together with
low signal-to-noise ratios. Moreover, part of the night-sky radiance is
related to extraterrestrial light sources and scattering inside the atmosphere
\citep[e.g.,][]{leinert98}. In particular, scattered moonlight, integrated and
scattered starlight, zodiacal light, and possible light pollution can be
significant sources of radiation. Hence, such components (which might be quite
uncertain) need to be subtracted to measure the nightglow continuum
\citep[e.g.,][]{sternberg72,noll12,trinh13}.

Despite the potential difficulties, \citet{barbier51} first noted a possible
continuum in the green wavelength range. In the subsequent decades, additional
constraints were found for a continuum in the visual wavelength range between
400 and 720\,\unit{nm}
\citep{davis65,broadfoot68,sternberg72,gadsden73,mcdade86}, where the density
of strong emission lines is relatively low. This continuum appeared to have a
flux of several Rayleigh per nanometre (\unit{R\,nm^{-1}}) with an increasing
trend towards longer wavelengths and a possible local maximum (or at least
plateau) near 600\,\unit{nm} \citep[e.g.,][]{gadsden73}. \citet{krassovsky51}
already proposed that this continuum could be produced by the reaction
\begin{reaction}\label{eq:NO+O}
  \mathrm{NO} + \mathrm{O} \rightarrow \mathrm{NO}_2 + h\nu.
\end{reaction}
The emission produced by this reaction, termed the \chem{NO_2} air afterglow,
was observed in laboratory discharge experiments and has a pressure-dependent
maximum, which is located around 580\,\unit{nm} for relevant atmospheric
densities \citep{fontijn64,becker72}. Space-based measurements of the emission
profile showed a peak between 90 to 95\,\unit{km}
\citep{savigny99,gattinger09,gattinger10,semenov14a}. First indicated by
ship-based latitude-dependent measurements \citep{davis65} and then studied in
more detail with the Optical Spectrograph and Infrared Imaging System (OSIRIS)
onboard the Odin satellite \citep{gattinger09,gattinger10}, the emission is
about an order of magnitude weaker at low latitudes compared with the polar
regions, where typical values near 580\,\unit{nm} are of the order of
10\,\unit{R\,nm^{-1}}.

However, \citet{evans10} found that an average OSIRIS spectrum for the low
latitude range from 0 to 40$^{\circ}$\,S did not match the expected spectral
distribution of the \chem{NO_2} air afterglow from Reaction~\ref{eq:NO+O}
because the data showed a more complex structure with a conspicuous relatively
narrow maximum near 600\,\unit{nm}. As an alternative explanation, they
proposed emission from electronically excited iron monoxide (\chem{FeO})
produced by
\begin{reaction}\label{eq:Fe+O3}
  \mathrm{Fe} + \mathrm{O}_3 \rightarrow \mathrm{FeO}^{\ast} + \mathrm{O}_2,
\end{reaction}
which had already been identified by \citet{jenniskens00} in the persistent
train of a Leonid meteor observed by an airborne optical spectrograph. Their
laboratory-based spectrum of these \chem{FeO} `orange arc' bands
\citep[see also,][]{west75,burgard06} also matched the OSIRIS spectrum quite
well. This interpretation implies that the low-latitude nightglow spectrum
around 600\,\unit{nm} can mainly be explained by a pseudo-contiuum consisting
of various ro-vibrational bands produced from the \chem{FeO} electronic
transitions $\mathrm{D}\,^5\Delta_i$ and $\mathrm{D^{\prime}}\,^5\Delta_i$ to
$\mathrm{X}\,^5\Delta_i$ \citep{cheung83,merer89,barnes95,gattinger11a}. Based
on the small OSIRIS data set covering five 24\,\unit{h} periods,
\citet{evans10} also found a good correlation of the pseudo-continuum and the
\chem{Na} chemiluminescence, which also depends on a reaction with ozone
(\chem{O_3}) and involves a chemical element supplied by the ablation of
cosmic dust \citep[e.g.,][]{plane15}. Covariations of \chem{Fe} and \chem{Na}
densities in the mesopause region were previously measured by lidar
\citep[e.g.,][]{kane93}. The corresponding results for the layer heights
of both metals also appear to agree well with the results from the OSIRIS
data suggesting a 3\,\unit{km} lower continuum emission layer with a peak at
about 87\,\unit{km}. The confidence in the \chem{FeO} scenario further
increased by the analysis of nine nights of sky radiance data obtained from
the Echelle Spectrograph and Imager (ESI) at the Keck\,II telescope on Mauna
Kea, Hawaii (20$^{\circ}$\,N) \citep{saran11}. The spectral range from 500 to
680\,\unit{nm} showed a structure with a peak at about 595\,\unit{nm}
consistent with laboratory data \citep{west75}. A slight shift of these (and
also the OSIRIS) data of about 5\,\unit{nm} towards longer wavelengths could
be explained by a higher effective vibrational excitation due to the low
frequency of quenching collisions at the lower pressures in the mesopause
region \citep{gattinger11a}. To date, the most detailed analysis of the shape
of the \chem{FeO} orange bands and their variability was reported by
\citet{unterguggenberger17}, based on 3,662 spectra of the \mbox{X-shooter}
echelle spectrograph \citep{vernet11} of the Very Large Telescope at Cerro
Paranal in Chile (24.6$^{\circ}$\,S, 70.4$^{\circ}$\,W). Clear seasonal
variations similar to those of the \chem{Na} nightglow, which were analysed in
the same study, were found. These variations could be characterised by a
combination of an annual and a semiannual oscillation (AO and SAO) with
relative amplitudes of 17 and 27\% and maxima in June/July and April/October,
respectively. Strong nocturnal trends were not observed. The spectrum (after
subtraction of other sky radiance components) appeared to have a stable
structure. The main peak between 580 and 610\,\unit{nm} with a mean intensity
of $23.2 \pm 1.1$\,\unit{R} contributed $3.3 \pm 0.8$\% to the total emission
in the range between 500 and 720\,\unit{nm}.

\citet{unterguggenberger17} did not see clear contributions of the reaction  
\begin{reaction}\label{eq:Ni+O3}
  \mathrm{Ni} + \mathrm{O}_3 \rightarrow \mathrm{NiO}^{\ast} + \mathrm{O}_2
\end{reaction}
with a bluer spectrum \citep{burgard06,gattinger11b}, i.e. with an expected
rise of the flux between 450 and 500\,\unit{nm} instead of around
550\,\unit{nm} as in the case of \chem{FeO}. This is in contrast to the
results for an average spectrum of the GLO-1 instrument on the Space Shuttle
mission STS\,53, where a ratio of the \chem{NiO} and \chem{FeO} intensities
integrated between 350 and 670\,\unit{nm} of $2.3 \pm 0.2$ was determined
\citep{evans11}. However, the same study also investigated OSIRIS mean spectra
of June/July over a period of three years, which resulted in much smaller
ratios of $0.3 \pm 0.1$, $0.1 \pm 0.1$, and $0.05 \pm 0.05$ that better agree
with \citet{unterguggenberger17}. \citet{evans11} also fitted the \chem{NO_2}
contribution from Reaction~\ref{eq:NO+O} relative to \chem{FeO} and found
ratios of 0.6, 0.2, and 0.0 with an uncertainty of 0.1. The correlation of
these ratios with those for \chem{NiO} and the extreme variation of the latter
suggest large uncertainties with respect to the impact of \chem{NiO}
nightglow.

At wavelengths slightly longer than 700\,\unit{nm}, early publications
indicated a significant increase of the radiance
\citep{broadfoot68,sternberg72,gadsden73}. However, the rocket-based
measurement of \citet{mcdade86} in Scotland (57$^{\circ}$\,N) only showed a
moderate radiance of 5.6\,\unit{R\,nm^{-1}} at 714\,\unit{nm} and
\citet{noxon78} measured an average of 7\,\unit{R\,nm^{-1}} at 857\,\unit{nm}
based on 15 nights at the Fritz Peak Observatory in Colorado (44$^{\circ}$\,N).
Low signal-to-noise ratios and the increasing strength of molecular
nightglow emission lines (\chem{OH} and \chem{O_2}) made measurements quite
challenging. The latter can also be seen in the shape of the nightglow
continuum of the Cerro Paranal sky model (25$^{\circ}$\,S) derived by
\citet{noll12}, based on 874 spectra of the FOcal Reducer and low dispersion
Spectrograph~1 (FORS\,1) covering a maximum wavelength range from 369 to
872\,\unit{nm}. While the region around the \chem{FeO} main peak (maximum of
about 6\,\unit{R\,nm^{-1}}) looks realistic, the steep rise at the longest
wavelengths is obviously related to the low resolving power of FORS\,1 of only
a few hundred.

At wavelengths above 900\,\unit{nm}, \citet{sobolev78} provided estimates of
about 9\,\unit{R\,nm^{-1}} at 927\,\unit{nm} and about 17\,\unit{R\,nm^{-1}} at
1,061\,\unit{nm} based on 5 nights of spectroscopic data from Zvenigorod,
Russia (57$^{\circ}$\,N). However, a flux of about 16\,\unit{R\,nm^{-1}} at
821\,\unit{nm} from the same study is distinctly higher than the result of
\citet{noxon78} for a similar wavelength. On the other hand, the Cerro Paranal
sky model provides for about 20\,\unit{R\,nm^{-1}} at 1,062\,\unit{nm}. In the
range between 1,032 and 1,775\,\unit{nm}, the continuum model was
coarsely derived from a small sample of 26 near-IR spectra from the relatively
new medium-resolution \mbox{X-shooter} spectrograph \citep{noll14}, where the
quality of the flux calibration and possible instrument-related continuum
contaminations were not yet known. In the set of considered wavelengths, the
residual continuum (after subtraction of other sky radiance components) shows
a minimum (for regions not affected by water vapour absorption) of about
9\,\unit{R\,nm^{-1}} at 1,238\,\unit{nm} and a maximum of about
87\,\unit{R\,nm^{-1}} at 1,521\,\unit{nm}. An increased flux level was also
measured by \citet{trinh13} with the Anglo-Australian Telescope in Australia
(31$^{\circ}$\,S) between 1,516 and 1,522\,\unit{nm}. For their sole continuum
window, they obtained $30 \pm 6$\,\unit{R\,nm^{-1}} based on 45 spectra with a
resolving power of 2,400, where strong \chem{OH} lines were suppressed by
means of fibre Bragg gratings \citep{ellis12}. The data of the covered five
nights also indicated a faster decrease of the continuum at the beginning of
the night than in the case of the \chem{OH} lines. \citet{maihara93} already
measured the range between 1,661 and 1,669\,\unit{nm} with a resolving power
of 1,900 in one night at Mauna Kea (20$^{\circ}$\,N) and found
$32 \pm 8$\,\unit{R\,nm^{-1}}. A similar flux of $36 \pm 11$\,\unit{R\,nm^{-1}}
was obtained by \citet{sullivan12} between 1,662 and 1,663\,\unit{nm} based on
the median of 105 spectra taken with a resolving power of 6,000 at Las
Campanas in Chile (29$^{\circ}$\,S). However, the Cerro Paranal sky model
provides here only about 13\,\unit{R\,nm^{-1}}. Moreover, 2\,\unit{h} of
observations with the GIANO spectrograph at the island La Palma (Spain,
29$^{\circ}$\,N) with the very high resolving power of 32,000 \citep{oliva15}
revealed a mean continuum level of about 16\,\unit{R\,nm^{-1}} in the range
from 1,519 to 1,761\,\unit{nm} avoiding regions affected by strong emission
lines. \citet{oliva15} also estimated that the presence of weak \chem{OH}
emission lines in the window used by \citet{maihara93} would require a
reduction of the radiance by 65\% resulting in about 11\,\unit{R\,nm^{-1}}.

The high uncertainties of the nightglow continuum in the near-IR made it
difficult to find explanations for the origin of the emission. The apparent
rise of the continuum beyond 700\,\unit{nm} led to the assumption that this
could be caused by another \chem{NO}-related reaction \citep{gadsden73}. As
derived by \citet{clough67} in the laboratory, the reaction 
\begin{reaction}\label{eq:NO+O3}
  \mathrm{NO} + \mathrm{O_3} \rightarrow \mathrm{NO}_2 + \mathrm{O_2} + h\nu
\end{reaction}
would be able to produce a broad continuum with a maximum near
1,200\,\unit{nm}. Later, \citet{kenner84} also investigated the reaction
\begin{reaction}\label{eq:NO+O3*}
  \mathrm{NO} + \mathrm{O^{\ast}_3} \rightarrow \mathrm{NO}_2 + \mathrm{O_2} +
  h\nu
\end{reaction}
involving excited \chem{O_3} with an emission maximum near 800\,\unit{nm}.
However, the increasing number of continuum measurements did not support a
large contribution from these reactions. Finally, calculations by
\citet{semenov14b} suggested that a radiance maximum of about
15\,\unit{R\,nm^{-1}} for Reaction~\ref{eq:NO+O} would lead to emission maxima
of about 5.4\,\unit{R\,nm^{-1}} for Reaction~\ref{eq:NO+O3} and about
0.3\,\unit{R\,nm^{-1}} for Reaction~\ref{eq:NO+O3*}, i.e. the reactions of
\chem{NO} with \chem{O_3} should only be minor contributions in the near-IR
especially at low latitudes, where the \chem{NO_2} air afterglow near
600\,\unit{nm} tends to be much weaker than given by \citet{semenov14b}. An
alternative proposal for a source of continuum emission was provided by
\citet{bates93}, who suggested metastable oxygen molecules that collide with
ambient gas molecules and then form complexes that dissociate by allowed
radiative transitions. However, there were no follow-up studies of this
scenario. Concerning laboratory measurements, \citet{bass52} and
\citet{west75} showed that \chem{FeO} does not only produce the orange bands.
Probably involving different electronic transitions, pseudo-contiuum emission
between 400 and 1,400\,\unit{nm} could be measured. It remains uncertain how
strong these additional bands could be under atmospheric conditions.

As there is obviously a lack of knowledge of the structure of the unresolved
nightglow emission and its variability (especially beyond the visual range),
we studied this topic by means of a large sample of well-calibrated
\mbox{X-shooter} spectra similar to those used by \citet{unterguggenberger17}
for \chem{FeO}-related research, i.e. mostly in the wavelength range between
560 and 720\,\unit{nm}. For the current study, we considered a much wider
wavelength range from about 300 to 1,800\,\unit{nm}. Moreover, the extended
data set covers 10 instead of 3.5 years, which allowed us to perform a more
detailed variability analysis. The data processing was also improved
\citep[cf.][]{noll22}. We discuss the data set, basic data processing, and
extraction of the nightglow (pseudo-)continuum in Sect.~\ref{sec:obs}. In
Sect.~\ref{sec:results}, we then describe the derivation of a mean continuum
spectrum, its decomposition into different components, the seasonal and
nocturnal variations of these components, the impact of the solar activity
cycle, and an estimate of the effective emission heights. As this analysis
revealed that it is necessary to introduce new nightglow emission processes,
we also explored several possible mechanisms for these emissions by carrying
out simulations with the Whole Atmosphere Community Climate Model (WACCM)
(Sect.~\ref{sec:modelling}). Finally, we draw our conclusions in
Sect.~\ref{sec:conclusions}.

\section{Observations}
\label{sec:obs}

\subsection{Data set}
\label{sec:dataset}

The \mbox{X-shooter} spectrograph \citep{vernet11} covers the wide wavelength
range between 300 and 2,480\,\unit{nm} with a resolving power between 3,200
and 18,400 depending on the arm (UVB: 300 to 560\,\unit{nm}, VIS: 550 to
1,020\,\unit{nm}, or NIR: 1,020 to 2,480\,\unit{nm}) and the variable width of
the entrance slit with a fixed projected length of 11$^{\prime\prime}$. For
standard slits with widths of 1.0$^{\prime\prime}$ (UVB), 0.9$^{\prime\prime}$
(VIS), and 0.9$^{\prime\prime}$ (NIR), the current nominal resolving power
amounts to about 5,400, 8,900, and 5,600, respectively. The entire
\mbox{X-shooter} data archive of the European Southern Observatory from the
start in October 2009 until September 2019 (i.e. 10 years of data) was
considered for this study. The NIR-arm data have already been used for
investigations focusing on \chem{OH} emission lines \citep{noll22,noll23}. As
described in these studies, the basic data processing was performed with
version v2.6.8 of the official reduction pipeline \citep{modigliani10} and
pre-processed calibration data. The resulting two-dimensional (2D)
wavelength-calibrated sky spectra were then reduced to one dimension (1D) by
averaging along the slit direction and adding possible sky remnants measured
in the 2D astronomical object spectrum extracted by the pipeline.

The flux calibration was performed by means of master response curves for
different time periods, which we derived from the comparison of
\mbox{X-shooter}-based spectra of spectrophotometric standard stars and the
theoretically expected spectral energy distributions \citep{moehler14}.
As discussed by \citet{noll22}, the NIR-arm spectra were calibrated by means
of 10 master response curves derived from data of the stars LTT\,3218 and
EG\,274, which have the highest fluxes in that wavelength regime. For the UVB
and VIS arms, more data of these stars and additional spectra of Feige\,110,
LTT\,7987, and GD\,71 \citep{moehler14} could be used due to the higher flux
at shorter wavelengths and the weaker disturbing nightglow emission. As this
increased the sample from 679 to 1,794 spectra and improved the
star-dependent time coverage, there were enough data to produce a series of 40
master response curves with a valid period of 3 months on average. This
allowed us to better correct the variability of the response, which tends to
increase towards shorter wavelengths due to the larger impact of dirt on the
mirrors. In the UVB arm at 370\,\unit{nm}, the individual response curves show
a relative standard deviation of about 9.1\%, whereas this percentage is only
about 3.5\% at 1,700\,\unit{nm}. From the flux-calibrated standard star
spectra, we obtain a residual variability of 3.6 and 1.7\% for the given UVB-
and NIR-related wavelengths. Uncertainties of about 2 to 3\% are typical for
most of the relevant wavelength range. A notable exception are wavelengths
around 560\,\unit{nm}, which are especially affected by the dichroic beam
splitting \citep{vernet11}. There, the flux variations amount to about 4 to
5\%. Finally, the absolute fluxes could show wavelength-dependent constant
systematic offsets of a few per cent as a comparison of the results for the
different standard stars indicate. We removed the differences by taking
LTT\,3218 as a reference. Hence, the absolute flux calibration depends on the
quality of the theoretical spectral energy distribution of this star
\citep{moehler14}.

Excluding very short exposures with less than 10\,\unit{s} and spectra with
very wide slits, which are mainly used for the spectrophotometric standard
stars, the final sample comprises about 56,000 UVB, 64,000 VIS, and 91,000 NIR
spectra. Although the three arms are usually operated in parallel, the numbers
differ due to arm-dependent splitting of observations. Failed processing is
another, albeit minor, issue. The exposure times can also be different. In
general, the sample is highly inhomogeneous due to different instrumental
set-ups, a wide range of exposure times up to 150\,\unit{min}, and different
possible residuals of the removed astronomical targets. Hence, the selection
of a high-quality sample for a specific research goal needs to be done very
carefully.

\subsection{Extraction of nightglow continuum}
\label{sec:extractcont}

\begin{figure*}[t]
\includegraphics[width=14cm]{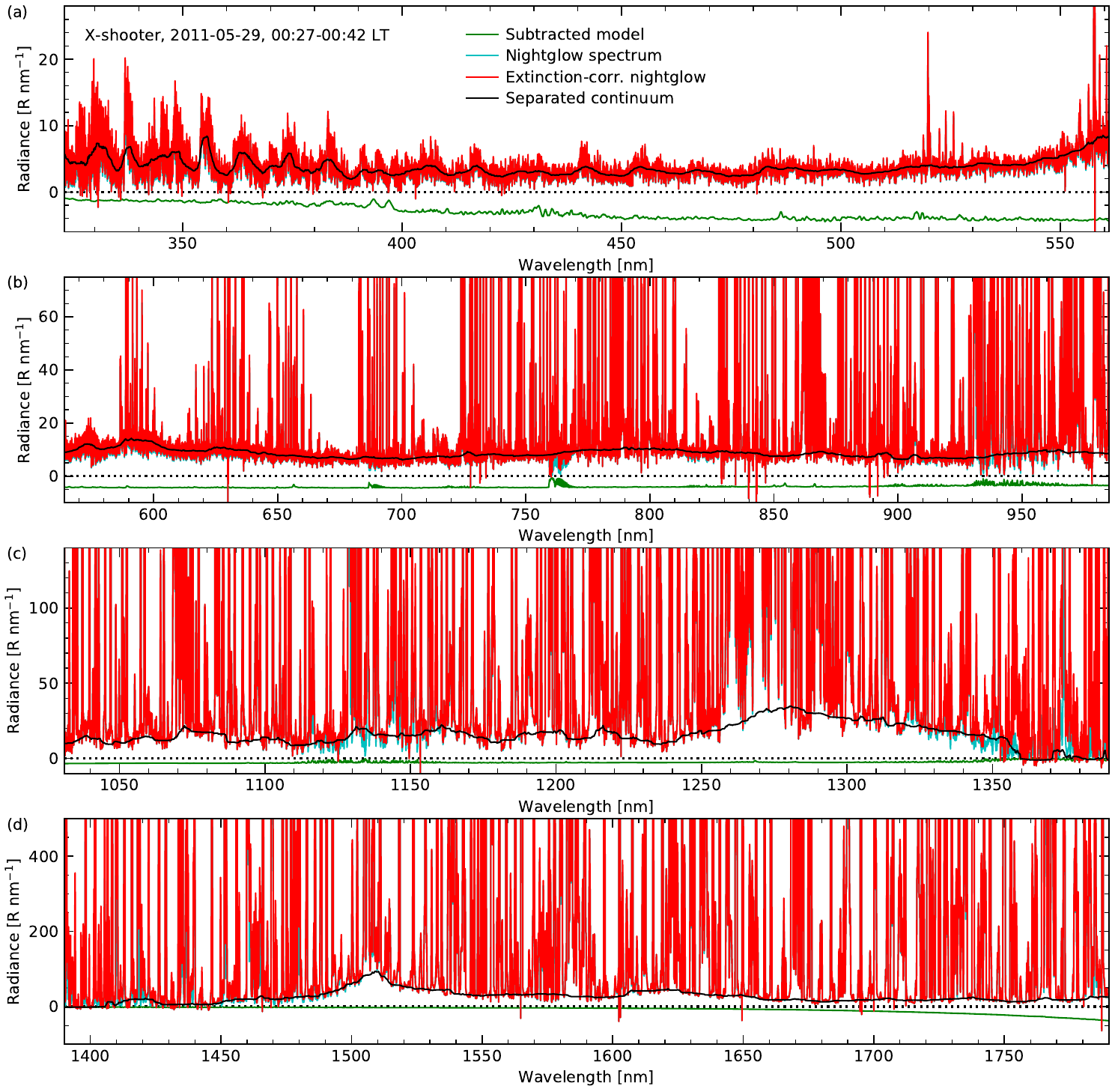}
\caption{Extraction of nightglow continuum for an \mbox{X-shooter} example
  spectrum with an exposure time of 15\,\unit{min} and standard width of the
  entrance slit for all three arms, i.e. UVB (a), VIS (b), and NIR (c and d).
  Wavelengths at the margins of the arm-related spectra and beyond
  1,790\,\unit{nm}, which are characterised by continuum data of low quality,
  are not shown. The green curve indicates the summed modelled contributions
  by scattered moonlight (not relevant here), zodiacal light, scattered
  starlight, and thermal emission of the telescope and instrument (multiplied
  by $-1$ for the plot) that were subtracted from the original sky spectrum.
  The cyan spectrum (or the overplotted red curve in the case of complete
  overlap) limited by the dotted zero line is the result of this subtraction
  and marks the nightglow emission. The red spectrum results from a
  continuum-optimised correction of the atmospheric extinction, i.e.
  absorption and scattering. The largest changes compared to the cyan curve
  are therefore related to wavelength ranges with strong absorption bands.
  Finally, the black solid curve shows the resulting nightglow continuum based
  on the application of percentile filters with wavelength-dependent
  percentile and width.}
\label{fig:sepcont}
\end{figure*} 

For the measurement of the \chem{OH} line intensities in the NIR arm
by \citet{noll22,noll23}, lines and underlying continuum were separated by
using percentile filters. For the present investigation of the nightglow
continuum, we applied the same approach to the other two arms
(Fig.~\ref{fig:sepcont}). As the density and strength of emission lines
depends on the wavelength, we used different combinations of percentile and
window width in order to optimise the separation. Concerning the percentile,
we applied a median filter in the UVB arm, a first quintile filter in the NIR
arm, and stepwise transition between both limiting percentiles in the VIS arm.
The window width for the major part of the spectral range was 0.8\% of the
central wavelength \citep[see also][]{noll22}. This width was further modified
primarily depending on the line density. In particular, extended relative
widths were applied to wavelengths affected by emission bands of \chem{O_2}
\citep[e.g.,][]{noll14,noll16} at 865\,\unit{nm} (0.02 instead of 0.008),
1,270\,\unit{nm} (0.04), and 1,580\,\unit{nm} (0.02). Nevertheless, remnants
of these bands could not be fully avoided (see Sect.~\ref{sec:meancont}).

Compared to the measurement of lines, the continuum separation was performed
after two preparatory steps. First, scattered moonlight, zodiacal light,
scattered starlight, and thermal emission of the telescope were calculated
using the Cerro Paranal sky model \citep{noll12,jones13} and subtracted from
the \mbox{X-shooter} spectra (Fig.~\ref{fig:sepcont}). Note that this is just
a rough correction with relatively high systematic uncertainties, especially
in the UVB arm when the Moon is up. On the other hand, the sky radiance
components related to direct or scattered light of sources from outside the
atmosphere are relatively weak in the NIR arm. In particular, around
1,500\,\unit{nm} the nightglow clearly dominates. However, the situation
deteriorates beyond 1,700\,\unit{nm}, where the non-zero emissivity of the
telescope and instrumental optical components leads to a rising thermal
continuum depending on the ambient temperature. The second preparatory step
was the correction of the atmospheric extinction by scattering and molecular
absorption. The former was performed by means of the recipes given by
\citet{noll12}, which consider the change of the reference Rayleigh and Mie
scattering from the sky model depending on the wavelength and zenith angle.
This correction is mostly relevant for the UVB arm, where flux changes by
several per cent are frequent, whereas the effect is negligible in the NIR
arm. Note that the nightglow brightness even tends to increase for spectra
taken close to the zenith due to Rayleigh scattering \citep{noll12}. Molecular
absorption especially by water vapour but also by \chem{O_3}, \chem{O_2},
\chem{CO_2}, and \chem{CH_4} reduces the detected radiance
\citep[e.g.,][]{smette15}. Here, we also used the sky model for a correction.
The continuum transmission curve was calculated for the given zenith distance,
given amount of precipitable water vapour (PWV), and otherwise standard
conditions at Cerro Paranal. For PWV values, we used the results from
\citet{noll22} based on intensity ratios of \chem{OH} lines in the NIR arm
with very different absorption fractions. The applied relations were
previously calibrated by means of local data from a Low Humidity And
Temperature PROfiler (\mbox{L-HATPRO}) microwave radiometer \citep{kerber12}.
Note that the simple division of a transmission curve does not provide correct
results for emission lines as their natural shape is not resolved. However, as
we are only interested in the continuum, we can neglect this issue here. As
long as the extinction is relatively small, the results of the correction are
reasonable. Nevertheless, nearly opaque wavelength regions, e.g. around
1,400\,\unit{nm} due to water vapour (Fig.~\ref{fig:sepcont}), cannot be
handled in this way. Even if the extinction was exactly known, small
uncertainties in the flux calibration and the modelled sky radiance components
would make a realistic correction impossible. Hence, the problematic
wavelength regions had to be excluded from the analysis.

After the subtraction of the line emission, the continuum spectra were
corrected for the increase of the emission with increasing zenith angle due to
a longer geometric path through the emission layer. This van Rhijn effect
\citep{vanrhijn21} was calculated assuming that the origin of the extracted
continuum was in the mesopause region. The results only weakly depend on
the reference height, which we set to 90\,\unit{km}. The validity of the
correction is supported by the consistent increase of the continuum flux with
increasing zenith angle in the whole wavelength regime for the optimised
sample described in Sect.~\ref{sec:meancont}.

\section{Results from observations}
\label{sec:results}

\subsection{The mean continuum}
\label{sec:meancont}

\begin{figure*}[t]
\includegraphics[width=17cm]{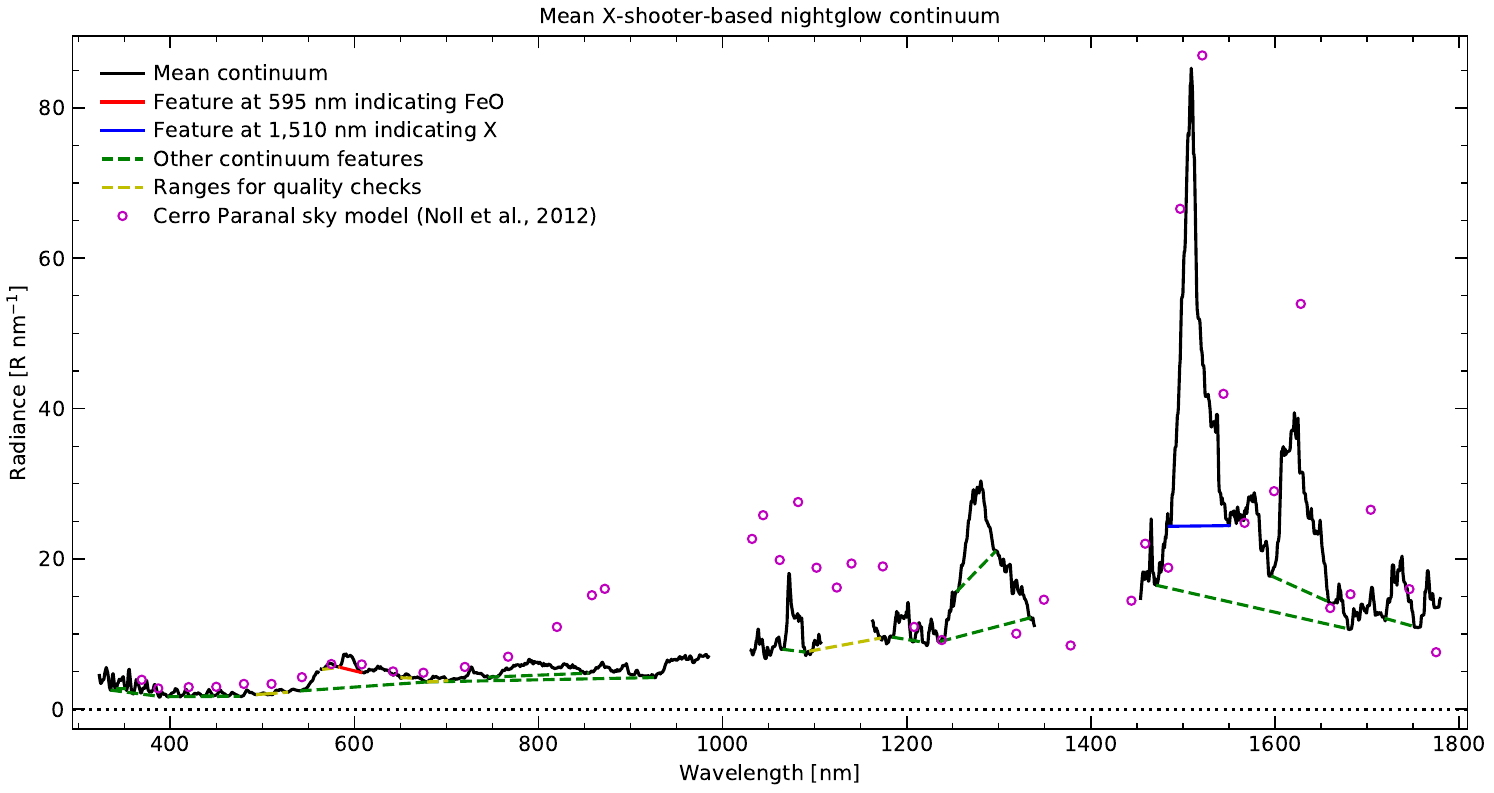}
\caption{Mean nightglow continuum spectrum at Cerro Paranal from 10,633
  combined \mbox{X-shooter} spectra. Wavelength ranges with systematic issues
  were not considered. The plot also shows the wavelength limits for different
  continuum features and their underlying continua that were used for the
  sample selection and the scientific analysis. The features centred on
  595\,\unit{nm} (red solid line) and 1,510\,\unit{nm} (blue solid line) are
  the main structures for the latter. Other reliable continuum features (or
  alternative definitions of their extent) are marked by green dashed lines.
  The ranges indicated by yellow dashed lines were only used for quality
  checks (including the detection of the contamination by astronomical
  objects). They do not mark real nightglow features. For a comparison, the
  open circles show the mean residual continuum of the Cerro Paranal sky model
  \citep{noll12}.}
\label{fig:meanfeat}
\end{figure*}

For the derivation of the mean nightglow continuum and the variability of the
continuum, we only selected the most reliable spectra. As a basic requirement,
data products of all three arms with similar temporal coverage had to be
available. In the case of arm-dependent differences in the number of exposures
(e.g., by shorter exposure times in the NIR arm than in the other arms), the
related spectra were averaged, weighted by the exposure time. The most
important selection criterion was the minimum exposure time, which was set to
10\,\unit{min} after several tests. The same cut was applied to the VIS-arm
sample studied by \citet{unterguggenberger17}. This criterion ensures that the
signal-to-noise ratio is high. However, the most important effect is the
reduction of continuum contamination by bright astronomical sources, which
tend to be observed with short exposure times. In order to keep the
non-nightglow sky radiance (and the uncertainties of its correction) low,
observations with the Moon above the horizon and an illumination of more than
50\% were excluded. In the end, these criteria led to 12,723 combined spectra,
which constitutes a substantial decrease compared to the full sample. In a
second selection procedure, various features in the continuum probably
belonging to the nightglow continuum, residuals of nightglow lines, or
residuals of astronomical objects (e.g. the H$\alpha$ line), and the remaining
underlying continuum were measured to identify spectra with suspected
artefactual contamination (Fig.~\ref{fig:meanfeat}). The resulting selection
limits (e.g. non-negative continuum fluxes), which were validated by visual
inspection of spectra with values close to the limits, led to a sample of
10,850 spectra. In a third step, the selection was further refined by the
search for abrupt changes in the times series of the continuum flux due to the
change of the astronomical target, which suggests a residual contamination.
Also validated by visual inspection, this procedure resulted in a final sample
of 10,633 combined spectra.

The mean of this data set is shown in Fig.~\ref{fig:meanfeat}. The spectrum
has gaps in wavelength ranges at the margins of the arms (due to high
systematic uncertainties) and strong atmospheric absorption (essentially by
water vapour). The latter explains the spectral upper limit at
1,780\,\unit{nm}, which also avoids wavelengths with strong thermal emission
of the telescope (Fig.~\ref{fig:sepcont}). At short wavelengths (i.e. in the
UVB arm), various bands related to the electronic upper states c, A$^{\prime}$,
and A of \chem{O_2} \citep[e.g.,][]{slanger03,cosby06} are visible. As the
bands are only partially resolved in the \mbox{X-shooter} spectra, the major
portion of the emission appears to be present as continuum.

The pronounced step in the continuum at about 555\,\unit{nm} and the peak at
about 595\,\unit{nm} indicates the presence of emission from the \chem{FeO}
orange bands
\citep{west75,jenniskens00,burgard06,evans10,saran11,gattinger11a,unterguggenberger17}.
The location of the step does not support significant contributions by NiO
\citep{burgard06,evans11,gattinger11b}, at least from the bluest systems
(\mbox{B-X} and \mbox{C-X}), which would already lead to a rise of the flux
below 500\,\unit{nm}. The shape of the continuum in this wavelength range also
excludes a significant contribution of \chem{NO_2} air afterglow
\citep{becker72,gattinger09,gattinger10,semenov14a}, which is not unexpected
as it is usually only bright at high latitudes (see also
Sect.~\ref{sec:intro}). Longwards of the peak at 595\,\unit{nm}, the continuum
shows only minor features in the VIS arm with a shallow local maximum at about
800\,\unit{nm}. There, the flux level is not higher than around the \chem{FeO}
main peak and lower than all published continuum measurements in this
wavelength range (Sect.~\ref{sec:intro}). At 857\,\unit{nm}, where
\citet{noxon78} obtained a relatively low value of about 7\,\unit{R\,nm^{-1}},
our mean flux is about 5.0\,\unit{R\,nm^{-1}}. For a comparison,
Fig.~\ref{fig:meanfeat} also shows the mean continuum from the Cerro Paranal
sky model of \citet{noll12}. While up to 770\,\unit{nm} the model continuum
is usually only slightly brighter than our \mbox{X-shooter}-based
measurements, the subsequent three data points are above
10\,\unit{R\,nm^{-1}}, which was most probably caused by the use of spectra
without sufficient resolving power.

In the NIR arm, our mean continuum is highly structured. In part, these
features are related to residuals of blends of strong \chem{OH} and
\chem{O_2} nightglow emission lines. In particular, remnants of the
\chem{O_2} bands at 1,270 and 1,580\,\unit{nm} related to the transitions
\mbox{(a-X)(0-0)} and \mbox{(a-X)(0-1)} can be identified
\citep[e.g.,][]{rousselot00,noll14,noll16}. Nevertheless, these features only
include a very small fraction of the total emissions, which were separated
with particularly wide filter windows because of the relatively high line
density (see Sect.~\ref{sec:extractcont}). The feature at about
1,080\,\unit{nm} is probably mainly related to the weak
\mbox{\chem{O_2}(a-X)(1-0)} band \citep[HITRAN database;][]{gordon22},
although the narrow maximum appears to be affected by \chem{OH} residuals. The
most striking continuum feature is certainly the high and narrow peak at about
1,510\,\unit{nm}. It is not related to residuals of strong lines. Hence, it is
probably composed of a high number of weak lines, which cannot be resolved
with the spectral resolving power of \mbox{X-shooter}. A feature with a
similar origin appears to be the peak at about 1,620\,\unit{nm}.

Both features do not appear to have been discussed previously in the airglow
literature. Nevertheless, they are already indicated in the coarse residual
continuum component of the Cerro Paranal sky model \citep{noll12}, which was
also derived from \mbox{X-shooter} spectra (see Sect.~\ref{sec:intro}).
Despite the high uncertainties in the model due to premature processing of
only a small number of spectra, the majority of the measurement points are
relatively close to our mean continuum. Notable exceptions in the NIR-arm
range are the fluxes at 1,628\,\unit{nm} (54\,\unit{R\,nm^{-1}}) and below
1,180\,\unit{nm}. Apart from possible problems with the separation of lines
and continuum, the offsets in the latter range suggest systematic issues with
the data processing. Data points in ranges that we excluded from our analysis
should be treated with caution. In Australia, \citet{trinh13} coincidentally
performed their continuum measurement of $30 \pm 6$\,\unit{R\,nm^{-1}} near
the emission peak between 1,516 to 1,522\,\unit{nm}. We find a higher flux of
about 50\,\unit{R\,nm^{-1}} for the same range. On the other hand, the mean
continuum between 1,661 and 1,669\,\unit{nm} in Fig.~\ref{fig:meanfeat}
amounts to about 14\,\unit{R\,nm^{-1}}, which is clearly lower than the
measurements of \citet{maihara93} and \citet{sullivan12}. However, it is
slightly brighter than a radiance of about 11\,\unit{R\,nm^{-1}} proposed by
\citet{oliva15} after the correction of the flux of \citet{maihara93} for the
contamination by faint \chem{OH} lines. Compared with the mean continuum flux
of about 16\,\unit{R\,nm^{-1}} obtained by \citet{oliva15} between 1,519 to
1,761\,\unit{nm} with high resolving power, our corresponding flux of about
22\,\unit{R\,nm^{-1}} is also slightly higher. Apart from differences in the
instrumental properties and the data processing, such discrepancies could also
be explained by the different observing sites and observing periods.
\citet{oliva15} only used 2\,\unit{h} of data taken at La Palma
(29$^{\circ}$\,N).

\subsection{Continuum decomposition}
\label{sec:contdecomp}

Most of the nightglow continuum emission in Fig.~\ref{fig:meanfeat} does not
exhibit clear features. In order to better understand this emission and its
relation to the identified features, we performed a decomposition of the
continuum in different components by means of the wavelength-dependent
variability pattern derived from the 10,633 selected spectra. Our approach was
to use non-negative matrix factorisation \citep[NMF; e.g.,][]{lee99,noll23} as
it is well suited for additive components without negative values. NMF
approximately decomposes an $m \times n$ matrix $\mathbf{X}$ without negative
elements into two non-negative matrices $\mathbf{A}$ and $\mathbf{B}$ with
sizes $m \times L$ and $L \times n$, respectively, by usually minimising the
squared Frobenius norm of $\mathbf{X} - \mathbf{A}\mathbf{B}$. For this
analysis, $m$, $n$, and $L$ are the number of wavelength positions, number of
spectra, and number of continuum components, respectively. As we sampled the
continuum spectrum with a resolution of 0.5\,\unit{nm} and only included the
ranges indicated in Fig.~\ref{fig:meanfeat}, $m$ was 2,479. For $L$, a
reasonable minimum is 4 since the features correlated with the \chem{FeO}
emission in the VIS arm, the unidentified features in the NIR arm, the
\chem{O_2} features in the UVB arm, and the residuals related to the
\mbox{\chem{O_2}(a-X)} bands in the NIR arm should be treated separately. This
definition of basic variability classes is supported by a check of the
correlations between the variability of the different measured features and
continuum windows. In the following, we call these classes \chem{FeO}(VIS),
\chem{X}(NIR), \chem{O_2}(UVB), and \chem{O_2}(NIR). The names refer to the
radiating molecule and location (in terms of the \mbox{X-shooter} arm) of the
main features of each class. It is not excluded that emission of other
molecules with a similar variability pattern can contribute. For the
application of the NMF, negative fluxes have to be avoided. Because of the
thorough sample selection procedure described above, the number of affected
data points was very small and negative values could therefore be replaced by
zeros without a significant change of the mean spectrum. Only between 1,031
and 1,037\,\unit{nm} (the shortest considered wavelengths in the NIR arm), the
mean flux increased by more than 1\%. For the derivation of the mean spectrum
of each component, we multiplied each of the resulting $L$ component spectra
consisting of $m$ data points with the mean of the $n$ corresponding scaling
factors. 

In the case of an application of the NMF with $L = 4$, it turned out that the
\chem{O_2} component in the UVB arm was not separated from the
\chem{FeO}-related features (similar to $L = 3$). This failure was probably
caused by the weakness of the \chem{O_2} features compared to the other
identified continuum structures. As a consequence, we increased the weight of
wavelength regions where a crucial feature was relatively strong by the
multiplication of suitable factors before the NMF and the division of the same
factors in the resulting component spectra. Consequently, the algorithm
minimised
$\left|\left| \mathbf{X^{\prime}} - \mathbf{A^{\prime}}\mathbf{B} \right|\right|^2_\mathrm{Fro}$,
where $\mathbf{X^{\prime}} = \mathbf{S}\mathbf{X}$ and
$\mathbf{A} = \mathbf{S}^{-1}\mathbf{A^{\prime}}$ with $\mathbf{S}$ being a
diagonal $m \times m$ matrix containing the wavelength-dependent scaling
factors. We tested different numbers and sizes of the windows. In the end, we
used 335 to 359\,\unit{nm}, 586 to 603\,\unit{nm}, 1,260 to 1,297\,\unit{nm},
and 1,497 to 1,521\,\unit{nm}, which maximised the weight of the main features
of the four variability classes. To find the best scaling factors, we defined
a cost function $C$ that uses the relative contributions $f_{ij}$ of the
$L = 4$ component spectra to the four corresponding feature windows as defined
above, i.e. we attempted to minimise
$C = 1 - \sum^L_{i=1} w_{i,\mathrm{max}} \, f_{i,\mathrm{max}}$, where
$w_{i,\mathrm{max}}$ and $f_{i,\mathrm{max}}$ being the weight and relative flux of
the most relevant feature window $j$ contributing to component $i$. Equal
weights of 0.25 for the four feature windows favoured solutions with
particularly large contributions of the two \chem{O_2}-related components.
However, the latter can be seen as contaminations of the \chem{FeO}(VIS) and
\chem{X}(NIR) components, which are obviously the primary targets of an
investigation of the nightglow continuum. Hence, we added the fractions with
different weights, finally choosing 0.33 for \chem{FeO}(VIS) and \chem{X}(NIR)
and 0.17 for \chem{O_2}(UVB) and \chem{O_2}(NIR). This (somewhat arbitrary but
non-critical) definition was sufficient to easily distinguish between
solutions with good component separation (best $C$ of 0.26) and those where
the separation failed, especially in the case of \chem{O_2}(UVB) and
\chem{FeO}(VIS) (best $C$ of 0.32). The relation between the scaling factors
in $\mathbf{S}$ and the structure of the component spectra turned out to be
complex. However, the variations within a certain class of solutions tended to
be relatively small. Hence, the solutions related to a satisfactory separation
of the four components as indicated by low $C$ are relatively robust. In any
case, there are two major components that dominate the visual and
near-infrared ranges.

In order to find minima of the cost function, we applied a simplicial homology
global optimisation \citep[SHGO;][]{endres18} algorithm in the ``sobol'' mode
with 512 sampling points and a limitation of the scaling factors between 1 and
200. The resulting list of local minima for $L = 4$ suggests an uncertainty in
the contribution fractions of several per cent for the windows in the UVB and
VIS arm and close to 1\% for the two windows in the NIR arm. Eventually, we
fine-tuned the most promising solution with scaling factors of about 139, 96,
68, and 65 (listed with increasing central wavelength of the feature window)
by starting an unconstrained simplex search algorithm \citep{nelder65} with
the given values as initial parameters. The resulting factors were about 1291,
865, 638, and 597, which differ from the initial values only by a nearly
constant factor. This points to a degeneracy of solutions, related to the fact
that the values are much higher than 1, i.e. the NMF results appear to be
mostly determined by the narrow feature windows. All reasonable local minima
found by SHGO in the parameter space are characterised by relatively high
values (limited to a maximum of 200), although the ratios of the four factors
can clearly differ.

\begin{figure*}[t]
\includegraphics[width=17cm]{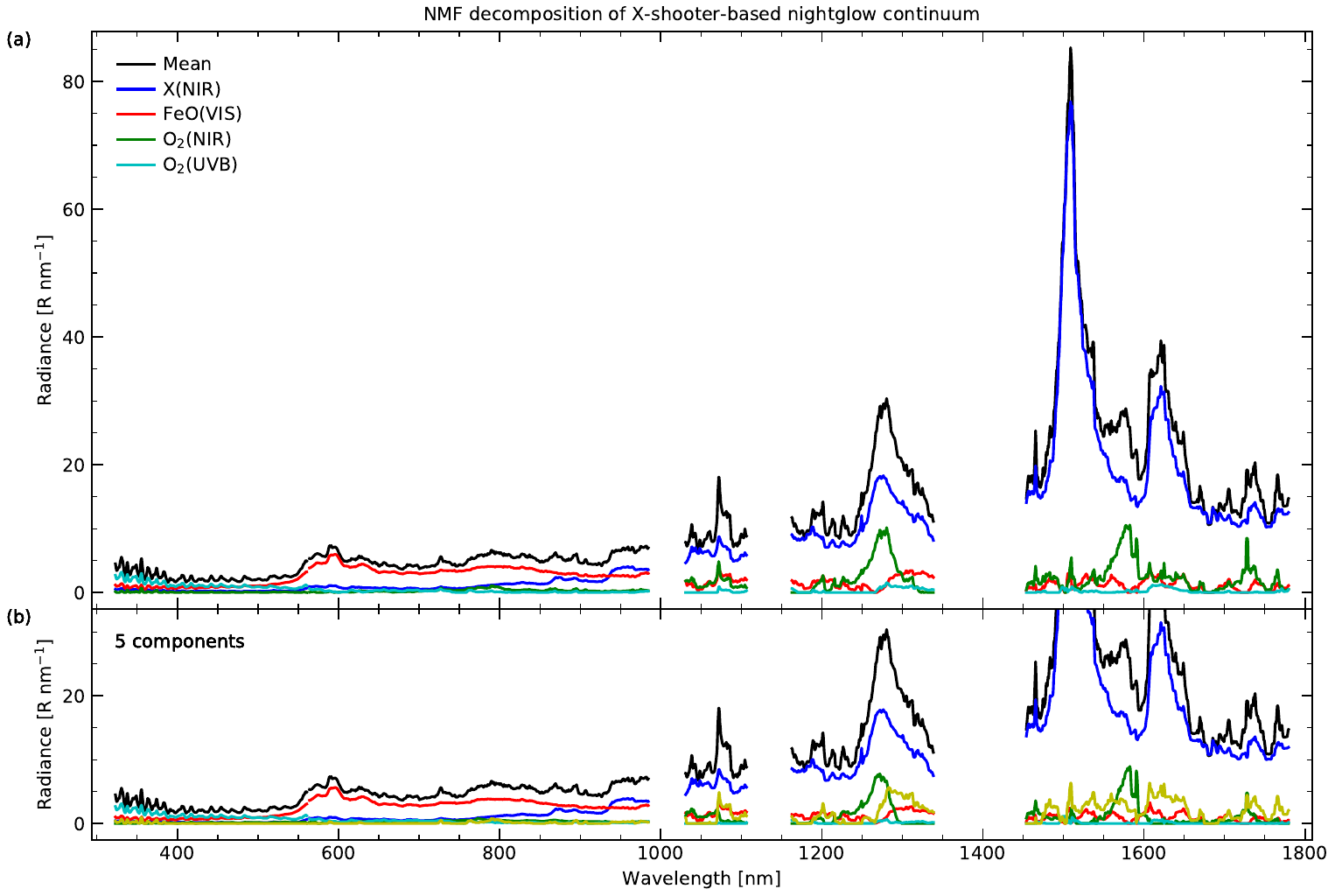}
\caption{Decomposition of mean nightglow continuum spectrum at Cerro Paranal
  (black curve) into (a) four and (b) five components by non-negative matrix
  factorisation of the selected 10,633 \mbox{X-shooter} spectra. The details
  of the procedure are discussed in Sect.~\ref{sec:contdecomp}. The four main
  components are labelled \chem{X}(NIR), \chem{FeO}(VIS), \chem{O_2}(NIR), and
  \chem{O_2}(UVB), which indicates the emitting molecule (if known) and the
  \mbox{X-shooter} arm with the dominating contribution. The fifth component
  in (b) appears to mainly consist of residuals of strong nightglow emission
  lines.}
\label{fig:contcomp}
\end{figure*}

The resulting mean continuum components based on refined simplex search are
shown in Fig.~\ref{fig:contcomp}a. The \chem{FeO}(VIS) and \chem{X}(NIR)
components contribute to the corresponding feature windows with 83.0\% and
95.1\%, respectively. Other reasonable solutions tend to show slightly lower
percentages. The dominance of these two components extends to wavelengths far
away from the main features. While \chem{FeO}(VIS) dominates almost the entire
VIS arm, \chem{X}(NIR) is the strongest mean component in the NIR arm. Similar
contributions appear to be present at the red end of the VIS arm. Below
500\,\unit{nm}, \chem{O_2}(UVB) becomes important with a dominating
contribution of 60.5\% in the reference range between 335 and 359\,\unit{nm}.
Nevertheless, \chem{FeO}(VIS) appears to still contribute with non-negligible
25.0\% there. In terms of the interpretation of this emission as based on
\chem{FeO}, this result is questionable as Reaction~\ref{eq:Fe+O3} should
only be exothermic by about 300\,\unit{kJ\,mol^{-1}} \citep{helmer94}, which
corresponds to a minimum wavelength of about 400\,\unit{nm}. Although the
separation of \chem{O_2}(UVB) and \chem{FeO}(VIS) shortwards of the \chem{FeO}
main peak seems to be the most uncertain result of the NMF-based continuum
decomposition, the \chem{FeO}(VIS) contributions in the UVB arm might support
the presence of the blue \chem{FeO} bands described by \citet{west75}. With a
higher significance, the high contribution of the component at about
800\,\unit{nm} might be explained by the presence of the \chem{FeO} IR bands
\citep{bass52,west75}, although the emission looks smoother than in the
laboratory, where it was not produced by Reaction~\ref{eq:Fe+O3}. According to
the analysis of \citet{gattinger11a}, the emission of the \chem{FeO} orange
bands is also less structured in the mesopause region than in the laboratory
due to a wider distribution of the vibrational populations. Moreover, it is
possible that residuals of other emissions in the \mbox{X-shooter} continuum
spectra led to an excessive removal of small-scale features. The direct
measurement of the broad feature between 745 and 855\,\unit{nm}
(Fig.~\ref{fig:meanfeat}) at least shows that the strength of this structure
is well correlated with the peak at 595\,\unit{nm}. The measurements in
the laboratory found \chem{FeO} emission up to 1,400\,\unit{nm}. The
\chem{FeO}(VIS) spectrum appears to show a similar extension. However, the
uncertainties of the minor contributions in the NIR arm compared to
\chem{X}(NIR) are large.

The \chem{FeO}(VIS) component could partly be produced by other metal-bearing
molecules if their emission showed a similar emission pattern. As already
discussed in Sect.~\ref{sec:meancont}, \chem{NiO} would be a candidate but the
shape of the continuum between 500 and 600\,\unit{nm} does not seem to allow a
major contribution. We searched for other possible molecules that could
produce a pseudo-continuum in the investigated wavelength regime. A survey of
the metal-related chemistry in the mesopause region turned out that another
abundant \chem{Fe}-containing reservoir species \citep{plane03,feng13,plane15}
could be a possible candidate. Unfortunately, chemiluminescence spectra of
these molecules do not appear to exist. Nevertheless, inspection of the
energetics of the relevant chemical reactions only left the reaction
\begin{reaction}\label{eq:FeOH+O3}
  \mathrm{FeOH} + \mathrm{O}_3 \rightarrow \mathrm{OFeOH}^{\ast} + \mathrm{O}_2
\end{reaction}
as sufficiently exothermic with up to 339\,\unit{kJ\,mol^{-1}}
(Sect.~\ref{sec:modelsetup}), i.e. almost the entire wavelength range in
Fig.~\ref{fig:contcomp} could be covered. We further discuss the possible role
of \chem{OFeOH} emission based on modelling results in
Sect.~\ref{sec:ressim}.

In Fig.~\ref{fig:contcomp}, the residuals of the strong \chem{O_2} bands at
1,270 and 1,580\,\unit{nm} are clearly identified by their dedicated component
\chem{O_2}(NIR). Nevertheless, the contributions are always smaller than those
related to \chem{X}(NIR). In the range between 1,260 and 1,297\,\unit{nm}, the
fraction is only 27.9\%. This percentage might be underestimated since
\chem{X}(NIR) shows a similar bump, which implies that the separation of both
components is incomplete. On the other hand, the structure in the mean
spectrum is broader than the \mbox{\chem{O_2}(a-X)(0-0)} band (especially at
longer wavelengths), which could suggest that at least a shallow
\chem{X}-related feature is present in this wavelength range. The weak
features at about 1,080 and 1,735\,\unit{nm} are not strong enough to be
classified with the NMF approach.

We checked how the components change if $L$ is set to 5 (keeping everything
else untouched). As indicated by Fig.~\ref{fig:contcomp}b, this modification
appears to mostly affect \chem{O_2}(NIR) by essentially reducing it to the
wavelengths of the two strong \chem{O_2} bands. The rest is mostly described
by the additional component, which seems to be sensitive to any other line
residuals (e.g. from \chem{OH}). Nevertheless, the version with $L = 4$ is
considered as the reference as it is more robust with respect to the important
\chem{FeO}(VIS) and \chem{X}(NIR) components, which are slightly weakened in
the case of $L = 5$. Tests with even larger numbers of components only showed
a higher complexity without improving the understanding of the nightglow
continuum.

\begin{table*}[th!]
\caption{Wavelength positions of \chem{HO_2} emission bands between 1,000 and
  1,800\,\unit{nm} observed in the laboratory in comparison to the
  \mbox{X-shooter}-based \chem{X}(NIR) spectrum}
\begin{tabular}{llccl}
\tophline
Upper state$^\mathrm{a}$ & Lower state$^\mathrm{a}$ & Peak$^\mathrm{b}$ &
Band origin$^\mathrm{c}$ & Presence in \chem{X}(NIR) \\
& & (\unit{nm}) & (\unit{nm}) & \\
\middlehline
$^2\mathrm{A}^{\prime}(002)$ & $^2\mathrm{A}^{\prime\prime}(000)$ &
1,130 & 1,130 & not measured (gap) \\
$^2\mathrm{A}^{\prime}(001)$ & $^2\mathrm{A}^{\prime\prime}(000)$ &
1,270 & 1,257 & moderate strength \\
$^2\mathrm{A}^{\prime}(002)$ & $^2\mathrm{A}^{\prime\prime}(001)$ &
(1,290) & 1,280 & possible but blended \\
$^2\mathrm{A}^{\prime}(000)$ & $^2\mathrm{A}^{\prime\prime}(000)$ &
1,430 & 1,423 & not measured (gap) \\
$^2\mathrm{A}^{\prime}(001)$ & $^2\mathrm{A}^{\prime\prime}(001)$ &
(1,480) & 1,446 & not clear (partly in gap) \\
$^2\mathrm{A}^{\prime\prime}(200)$ & $^2\mathrm{A}^{\prime\prime}(000)$ &
1,510 & 1,505 & very strong \\
$^2\mathrm{A}^{\prime}(000)$ & $^2\mathrm{A}^{\prime\prime}(001)$ &
1,690 & 1,670 & no clear feature \\
$^2\mathrm{A}^{\prime}(001)$ & $^2\mathrm{A}^{\prime\prime}(002)$ &
1,730 & & weak feature \\
\bottomhline
\end{tabular}
\belowtable{
\begin{list}{}{}
\item[$^\mathrm{a}$] electronic and vibrational ($v_1v_2v_3$) levels
\item[$^\mathrm{b}$] as given by \citet{becker74} for low-resolution data
  (unresolved bands with calculated wavelengths in parentheses)
\item[$^\mathrm{c}$] as measured by \citet{becker78} and/or \citet{tuckett79}
  at medium/high resolution
\end{list}
}
\label{tab:HO2peaks}
\end{table*}

If there is only one chemical process that produces the \chem{X}(NIR)
spectrum, the reaction that produces the excited states needs to be
sufficiently exothermic to explain the derived emission at least between about
900 and 1,800\,\unit{nm}. The solution might be a molecule like \chem{OFeOH},
where the variability pattern could also be quite different from the
\chem{FeO} emission variations. It is also possible that the radiating
molecule does not include a metal atom if it is sufficiently complex to be
suitable to produce a pseudo-continuum in a wide wavelength range. Here,
the hydroperoxyl radical (\chem{HO_2}) appears to be the best candidate.
\chem{HO_2} is often discussed in terms of mesospheric chemistry with respect
to the reaction
\begin{reaction}\label{eq:HO2+O}
  \mathrm{HO}_2 + \mathrm{O} \rightarrow \mathrm{OH}^{\ast} + \mathrm{O}_2,
\end{reaction}
which is an alternative production mechanism for vibrationally-excited
\chem{OH} \citep[e.g.,][]{makhlouf95,xu12,panka21}. The latest results of
\citet{panka21} suggest that this pathway contributes significantly to the
concentration of \chem{OH} in the lower mesopause region around 80\,\unit{km},
although the resulting vibrational level distribution remains uncertain.
The abundance of \chem{HO_2} in the mesosphere has been observed from the
ground \citep{clancy94,sandor98} and from space
\citep{pickett08,baron09,kreyling13,millan15} based on individual lines in the
microwave range. While the highest daytime densities tend to be between 75 and
80\,\unit{km}, the weaker nighttime maxima were observed between 80
and 90\,\unit{km} at low latitudes, with the highest altitudes before sunrise
\citep{kreyling13}. The near-IR spectrum of \chem{HO_2} has been
widely investigated in the laboratory
\citep[e.g.,][]{hunziker74,becker74,becker78,tuckett79,holstein83,fink97}.
Emission was mainly produced by the reaction
\begin{reaction}\label{eq:HO2+O2a}
  \mathrm{HO}_2 + \mathrm{O}_2(\mathrm{a}^1\Delta_{\mathrm{g}}) \rightarrow
  \mathrm{HO}_2^{\ast} + \mathrm{O}_2.
\end{reaction}
The resulting bands up to 1,800\,\unit{nm} listed by \citet{becker74} are
given by Table~\ref{tab:HO2peaks}. The peak wavelengths are complimented by
band origins derived from higher-resolution data of \citet{becker78} and
\citet{tuckett79}. In some cases, the provided wavelengths were obtained from
the combination of the molecular data of both publications. Most bands in
Table~\ref{tab:HO2peaks} are related to transitions between the lowest-lying
excited electronic state $^2\mathrm{A}^{\prime}$ and the ground state 
$^2\mathrm{A}^{\prime\prime}$ that involve the $v_3$ \chem{O-OH} stretching
vibration of both levels. Interestingly, the excitation energies of
$^2\mathrm{A}^{\prime}$(001) and \chem{O_2}(a$^1\Delta_{\mathrm{g}}$) are almost
identical. As a consequence, the resulting near-resonant energy transfer
produces the \chem{HO_2} emission feature near 1,270\,\unit{nm}. This is
appealing as this would explain our NMF results in this wavelength region. The
strongest band in the experiments cannot be checked as wavelengths around
1,430\,\unit{nm} corresponding to the \mbox{(000-000)} band were excluded in
our analysis due to the strong absorption by atmospheric water vapour (see
Fig.~\ref{fig:sepcont}). However, the most promising argument for \chem{HO_2}
as \chem{X} is the only purely vibrational band in the list. The
\mbox{(200-000)} transition that involves the \chem{OO-H} stretching mode
peaks near 1,500 and 1,510\,nm \citep[e.g.,][]{hunziker74}. The second maximum
clearly agrees with the peak of our \chem{X}(NIR) main feature. The
invisibility of the first maximum might be caused by systematic uncertainties
in the continuum separation near the Q branch of \mbox{\chem{OH}(3-1)}
(Fig.~\ref{fig:sepcont}) combined with a less pronounced dip at the band
origin in the nightglow spectrum.

Other bands of Table~\ref{tab:HO2peaks} that can be checked should peak near
1,690 and 1,730\,\unit{nm}. While we see a possible weak feature in the
\mbox{X-shooter} spectrum in the latter case, there is no clear structure near
1,690\,\unit{nm}. This result is not necessarily an argument against
\chem{HO_2} as the vibronic \mbox{(000-001)} band was much weaker than the
\mbox{(001-000)} band near 1,270\,\unit{nm} in the experiment of
\citet{fink97}. A more crucial issue could be the missing evidence for a
strong feature near 1,620\,\unit{nm} (Fig.~\ref{fig:contcomp}) in the
laboratory. If \chem{HO_2} is indeed the correct emitter (i.e. species
\chem{X}), then the population distributions need to be very different in the
mesopause region, where the pressure is much lower (3 orders of magnitude)
compared to the experiment of \citet{fink97}. The spectrum of the latter study
that covers the wavelength range between 1,200 and 1,800\,\unit{nm} indicates
weaker emission at 1,510\,\unit{nm} than at 1,270\,\unit{nm}. This could point
to an increased importance of purely vibrational transitions in the nightglow.
Various additional bands might be visible, which could explain the
1,620\,\unit{nm} feature and the relatively high emission over a wide
wavelength range. In contrast to \chem{X}(NIR), the laboratory spectrum does
not show significant emission between 1,320 and 1,350\,\unit{nm} as well as
below 1,200\,\unit{nm}. The latter is certainly related to
Reaction~\ref{eq:HO2+O2a}, which limits the emission below 1,270\,\unit{nm}.
Nevertheless, \citet{becker74} could measure the vibronic \mbox{(002-000)}
band near 1,130\,\unit{nm} (in a gap in Fig.~\ref{fig:contcomp}) and explained
it by already vibrationally excited \chem{HO_2} as reaction partner. In a
similar way, \citet{holstein83} assumed that two subsequent collisions with
\chem{O_2}(a$^1\Delta_{\mathrm{g}}$) are required to excite this band and
additional weaker bands in the range between 800 and 1,100\,\unit{nm} that
involve $^2\mathrm{A}^{\prime}$ $v_3$ states between 3 and 6. The lower
wavelength limit for the observed emission would be consistent with the shape
of the \chem{X}(NIR) component.

Importantly, \citet{holstein83} found that chemiluminescence can also be
generated at wavelengths longer than 800\,\unit{nm} by the main atmospheric
production process of \chem{HO_2}
\citep[e.g.,][]{makhlouf95}
\begin{reaction}\label{eq:H+O2+M}
  \mathrm{H} + \mathrm{O}_2 + \mathrm{M} \rightarrow \mathrm{HO}_2^{\ast} +
  \mathrm{M}
\end{reaction}
with \chem{M} being an arbitrary collision partner (i.e. \chem{N_2} and
\chem{O_2} in the mesosphere). Here, the spectrum showed a weaker dependence
of the intensities of the vibronic \mbox{(00$v_3^{\prime}$-000)} bands on
$v_3^{\prime}$ than in the case of collisions with
\chem{O_2}(a$^1\Delta_{\mathrm{g}}$). The recombination of \chem{H} and
\chem{O_2} is also sufficiently exothermic to produce emission potentially as
far as about 600\,\unit{nm}. Other chemical reactions producing excited
\chem{HO_2} could also play a role (see Sect.~\ref{sec:modelling}). In the
view of the remaining uncertainties, we do not replace \chem{X} by a specific
molecule in the following. First, further properties of the unknown emission
have to be discussed.

\subsection{Intensity climatologies}
\label{sec:intclim}

The NMF also returns the scaling factors of each component for each input
spectrum. The resulting variability patterns are the basis for the separation
of the components shown in Fig.~\ref{fig:contcomp}. Before we discuss the
variations of the different components, we focus on a comparison of the
variability of the two most interesting, directly measured features. These are
the peaks at about 595 and 1,510\,\unit{nm}, which are closely related to the
NMF components \chem{FeO}(VIS) and \chem{X}(NIR). The two peaks were measured
by the interpolation between 584 and 607\,\unit{nm} as well as 1,485 and
1,550\,\unit{nm} for the derivation of the underlying continuum (see
Fig.~\ref{fig:meanfeat}). The latter feature was then subtracted from the
integrated flux in the same wavelength intervals in order to obtain the
feature intensity. \citet{unterguggenberger17} already measured the \chem{FeO}
main peak with a similar approach using 3,662 \mbox{X-shooter} VIS-arm spectra
taken between October 2009 and March 2013. The continuum spectra were
extracted slightly differently by interpolating between wavelengths
significantly affected by line emission and leaving the rest of the spectrum
untouched. As that method causes noisier spectra than in the case of the
percentile filters used in this study (45th percentile and a relative width of
0.008 of the filter at the peak-related wavelengths), the positions for the
interpolation on both sides of the peak were adapted to the corresponding flux
minima in each spectrum. \citet{unterguggenberger17} reported a reference
intensity of the \chem{FeO} main peak based on a harmonic model of the
seasonal variations of $23.2 \pm 1.1$\,\unit{R}. Our sample shows a mean of
27.0\,\unit{R}, which indicates good agreement under consideration of the
differences in the sample and the measurement approaches. For comparison, the
mean of the peak at 1,510\,\unit{nm} amounts to 1,371\,\unit{R}, i.e. it is
about 51 times stronger. The ratio would be even higher for wider feature
limits around 1,510\,\unit{nm} that would be reasonable for the \chem{X}(NIR)
component in Fig.~\ref{fig:contcomp}. For example, the interval between 1,472
and 1,591\,\unit{nm} would lead to 1,983\,\unit{R}, i.e. a rise by a factor of
1.45 compared to the tighter interval defined in Fig.~\ref{fig:meanfeat},
which we preferred for the measurements in the full spectra in order to avoid
the varying contamination by the residuals of the \mbox{\chem{O_2}(a-X)(0-1)}
band.

For the study of the variability, we calculated 2D climatologies of local time
and day of year in the same way as described in \citet{noll23} for \chem{OH}
emission lines. The measured \chem{OH} line intensities were not directly used
\citep[see also][]{noll22}. Instead, the time series were divided into bins of
30\,\unit{min} and intensities of data with central times in a certain bin
were averaged weighted by the exposure time. The reason for this approach was
the wide range of exposure times down to 10\,\unit{s}, which could lead to a
high weight of a large number of short low-quality exposures (partly clustered
in time) in the resulting climatologies if the individual measurements were
used. For the NMF-related sample of this study, this is less problematic as
only exposures with a minimum length of 10\,\unit{min} were considered.
Nevertheless, we also performed this preparatory step for the sake of
consistency. \citet{noll23} only selected those bins with a minimum filling of
10\,\unit{min}. This criterion is automatically fulfilled by the NMF-related
sample. However, this approach led to a reduction of the number of data points
from 10,633 to 7,971 (75.0\%).

The climatologies consist of a grid of the centres of the 12 hours between
18:00 and 06:00 LT (the local time related to the solar mean time at Cerro
Paranal) and the centres of the 12 months in days of year. The reference
values for these grid points were derived from the average of all bins within
a radius of 1\,\unit{h} and 1 average month at least if a minimum of 200 bins
were selected. In the case of fewer bins, the radius was increased in steps of
0.1 until the criterion was fulfilled. As this issue mainly concerns grid
points close to twilight, the temporal resolution at the margins of the
climatologies is lower than in the middle of the night. In the LT range
between 20:00 and 04:00, the mean relative radius was 1.08. The final
climatologies are provided relative to the effective mean, for which the grid
point data were averaged weighted by the night contribution (defined by a
minimum solar zenith angle of 100$^{\circ}$) of the surrounding cells.
Moreover, they are given for a reference solar radio flux at 10.7\,\unit{cm}
\citep{tapping13} averaged for 27 days of 100 solar flux units (\unit{sfu}).
This approach compensates for values between 88 and 110\,\unit{sfu} (with an
effective value of 99\,\unit{sfu}) for the different grid points assuming a
linear relation between the investigated property and the solar radio flux.
The corrections are of the order of a few per cent at most. Hence, the
uncertainties in the regression results do not critically affect the quality
of the climatologies. The effective intensities of the two features derived
from the final climatologies are 27.3 and 1,386\,\unit{R}, which are very
close to the mean values for the individual measurements.

In order to better understand the quality of the climatologies, we also
calculated them for a minimum sample size of 400 for each grid point as this
was the limit used by \citet{noll23} for a total number of bins of up to
19,570. As the NMF-related data set is distinctly smaller, this choice causes
smoother climatologies due to the necessary increase of the selection radius.
Between 20:00 and 04:00 LT, its mean is 1.43. On the other hand, larger
subsamples can reduce the statistical uncertainties. Despite these
differences, the intensity climatologies look very similar. The correlation
coefficients for the comparison of the versions with lower limits of 200 and
400 bins (only considering grid cells with a nighttime fraction higher than
20\%) for the two features are higher than $+0.98$. The impact is larger on
the climatologies of the solar cycle effect (SCE), i.e. the relations between
the investigated property and the solar radio flux. For this comparison, the
coefficients are $+0.86$ and $+0.80$ for the features at 595 and
1,510\,\unit{nm}.

As another test, we investigated the impact of the increase of the total
sample size on the climatologies. For the two continuum features, the data
selection can be extended as it is only required that they can be measured
satisfactorily irrespective of the situation at other wavelengths. As the
feature at 1,510\,\unit{nm} is relatively bright, the number of suitable
spectra could be increased to 45,037 including data with minimum exposure
times of 3\,\unit{min} (instead of 10\,\unit{min}). This sample resulted in
17,482 30\,\unit{min} bins (an increase by a factor of 2.2), which allowed us
to calculate an intensity climatology with a minimum subsample size of 400
without resolution losses. The result correlates very well with the
climatology of the small sample with high resolution. The correlation
coefficient $r$ is $+0.996$. On the other hand, the SCE-related climatology
indicates an $r$ of only $+0.38$, probably partly caused by a vanished outlier
in the case of the large sample. Hence, the details of the SCE with respect to
LT and day of year remain uncertain, whereas the intensity-related results
appear to be quite robust.

In the case of the \chem{FeO} main peak, the extension of the data set was
more limited as the feature is distinctly fainter and the sample of VIS-arm
spectra is smaller. Finally, we selected 22,322 intensity measurements with a
minimum exposure time of 5\,\unit{min}, which were converted into 12,785 bins
corresponding to an increase of the sample size by a factor of 1.6. The
climatology was then also calculated using a minimum subsample size of 400.
The resulting intensity variations show a high similarity with those of the
small sample as an $r$ of $+0.986$ indicates. Nevertheless, there appears to
be an issue with the large sample with respect to the effective intensity of
the climatology, which turned out to be 11.5\% higher than in the case of the
small sample. The effective intensity of the 1,510\,\unit{nm} feature only
increased by 2.3\%. This points to a significant contamination by remnants
especially of astronomical objects, suggesting that a relaxation of the
selection criteria is problematic for the 595\,\unit{nm} feature.
Interestingly, the correlation coefficient for the SCE and the 595\,\unit{nm}
feature is $+0.76$, i.e. it is higher than for the NIR-arm feature. This could
be related to a smoother climatology without clear outliers. 

\begin{figure*}[t]
\includegraphics[width=17cm]{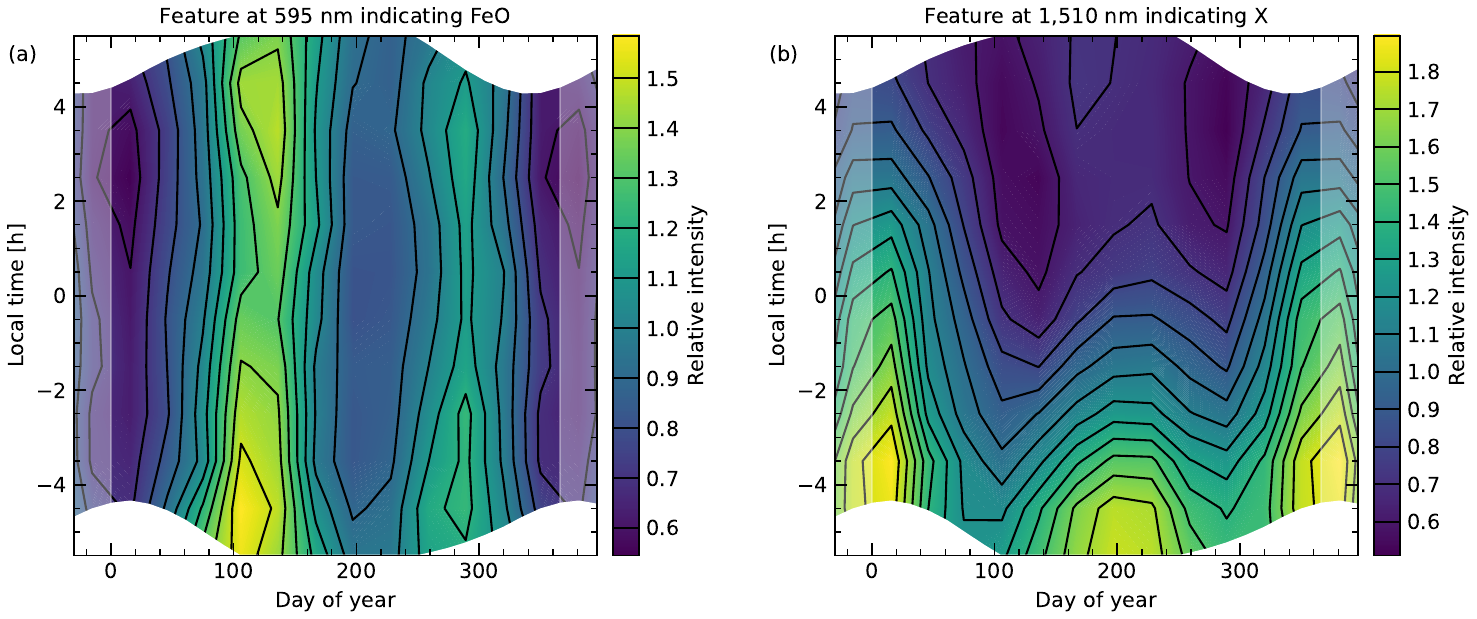}
\caption{Climatologies of intensity relative to the mean as a function of
  local time (step size of 1\,\unit{h}) and day of year (step size of 1 month)
  for the continuum features at (a) 595\,\unit{nm} and (b) 1,510\,\unit{nm}
  based on a sample of 7,971 30\,\unit{min} bins and a minimum subsample size
  of 200. The climatologies are representative of a solar radio flux of
  100\,\unit{sfu}. The coloured contours are limited to times with solar
  zenith angles larger than 100$^{\circ}$. Lighter colours at the left and
  right margins mark repeated parts of the variability pattern.}
\label{fig:featclim_I}
\end{figure*}

As illustrated by the previous discussion, the 2D climatologies of the
relative intensity variations of the 595 and 1,510\,\unit{nm} features can be
considered as robust. For the NMF-related sample and a minimum subsample size
of 200, these climatologies are shown in Fig.~\ref{fig:featclim_I}. The
variations of the \chem{FeO} emission peak in (a) are mainly characterised by
a semiannual oscillation (SAO) with maxima in April/May (nightly averaged
relative intensity of 1.40) and October (1.17) and minima in January (0.61)
and July/August (0.86). The higher intensities for the maxima and minima in
April/May and July/August also indicate an annual oscillation (AO) with a
maximum in austral autumn/winter. This result is in good agreement with the
harmonic fits of the smaller \mbox{X-shooter} data set of
\citet{unterguggenberger17} (see Sect.~\ref{sec:intro}), which only included
spectra until March 2013. WACCM simulations \citep{feng13} suggest that the AO
is mainly driven by the \chem{Fe} concentration, which depends on the meteoric
injection rate (maximum in March/April) and subsequent chemical reactions,
whereas the SAO is mainly linked to the intra-annual variations of the other
\chem{FeO}-producing reactant, i.e. \chem{O_3}. The concentration maxima of
the latter shortly after the equinoxes can also be seen in data of the
Sounding of the Atmosphere using Broadband Emission Radiometry (SABER)
instrument onboard the Thermosphere Ionosphere Mesosphere Energetics Dynamics
(TIMED) satellite \citep{russell99} for Cerro Paranal at 89\,\unit{km}
analysed by \citet{noll19}. \citet{unterguggenberger17} also investigated the
average nocturnal patterns in the different seasons and found only weak
changes without clear trend. With the larger sample of this study, these
observations can be confirmed. The changes do not exceed 10 to 20\% of the
mean value in most parts of the climatology. On average, there is a shallow
minimum in the middle of the night. The month-dependent nocturnal variations
could be related to the impact of tides. The corresponding features are
visible more clearly in the \chem{O} number density at about 89\,\unit{km}
\citep{noll19} and \chem{OH} emissions especially of lines with relatively
high rotational quantum number \citep{noll23}, which are not particularly
affected by the rapid nocturnal loss of daytime-produced \chem{O} close to
80\,\unit{km}. 

The 2D climatology of the continuum feature at 1,510\,\unit{nm} in
Fig.~\ref{fig:featclim_I}b is very different from the pattern observed for the
structure at 595\,\unit{nm}. There is a striking decrease of the intensity by
a factor of 2 to 3 from the beginning to the end of the night in all months of
the year. Only in the middle of the year in the morning, a plateau appears to
be reached. This pattern points to a loss of the excited radiating molecules
with the start of the night, i.e. the chemical set-up appears to be different
between day and night. Examples of such cases are \chem{OH} emission
especially below 84\,\unit{km} \citep{marsh06,noll23} due to cessation of 
\chem{O_2} photolysis, and \mbox{\chem{O_2}(a-X)} emission \citep{noll16}
due to the cessation of \chem{O_3} photolysis. Interestingly, \citet{trinh13}
previously reported a decrease of the continuum between 1,516 and
1,522\,\unit{nm} in the first half of the night based on spectra from the
Anglo-Australian Telescope (31$^{\circ}$\,S) taken during five nights in
September 2011 (see Sect.~\ref{sec:intro}). The decrease in the evening
appeared to be slightly faster than in the case of the Q branch of
\mbox{\chem{OH}(3-1)}. This is consistent with our results from a comparison
with the corresponding \chem{OH} line climatologies from \citet{noll23}, which
indicated an about 15\% higher intensity reduction between 19:30 and 21:30 LT
for the continuum peak on average. The seasonal variations of the
1,510\,\unit{nm} feature show a main maximum in January (nightly averaged
relative intensity of 1.59), a secondary maximum in July/August (1.02), and
minima in April (0.77) and October (0.84). This behaviour is almost the exact
opposite of the seasonal variations of the \chem{FeO} main peak. The
correlation coefficient for the monthly mean values is $-0.90$. This
anticorrelation does not seem to support a strong impact of \chem{O_3} in the
production of emitter \chem{X}. This can be an issue for \chem{OFeOH} as
produced by Reaction~\ref{eq:FeOH+O3}. On the other hand, the seasonal
variability of the 1,510\,\unit{nm} emission is reminiscent of the one
expected for atomic hydrogen (\chem{H}) \citep{mlynczak14}, which could be an
argument for the participation of \chem{H} in the production of the
radiating molecule. Interestingly, this is fulfilled by the \chem{HO_2}
production process given in Reaction~\ref{eq:H+O2+M}. Based on our WACCM
simulations, we discuss this topic in Sect.~\ref{sec:ressim} in more detail.

\begin{figure*}[t]
\includegraphics[width=17cm]{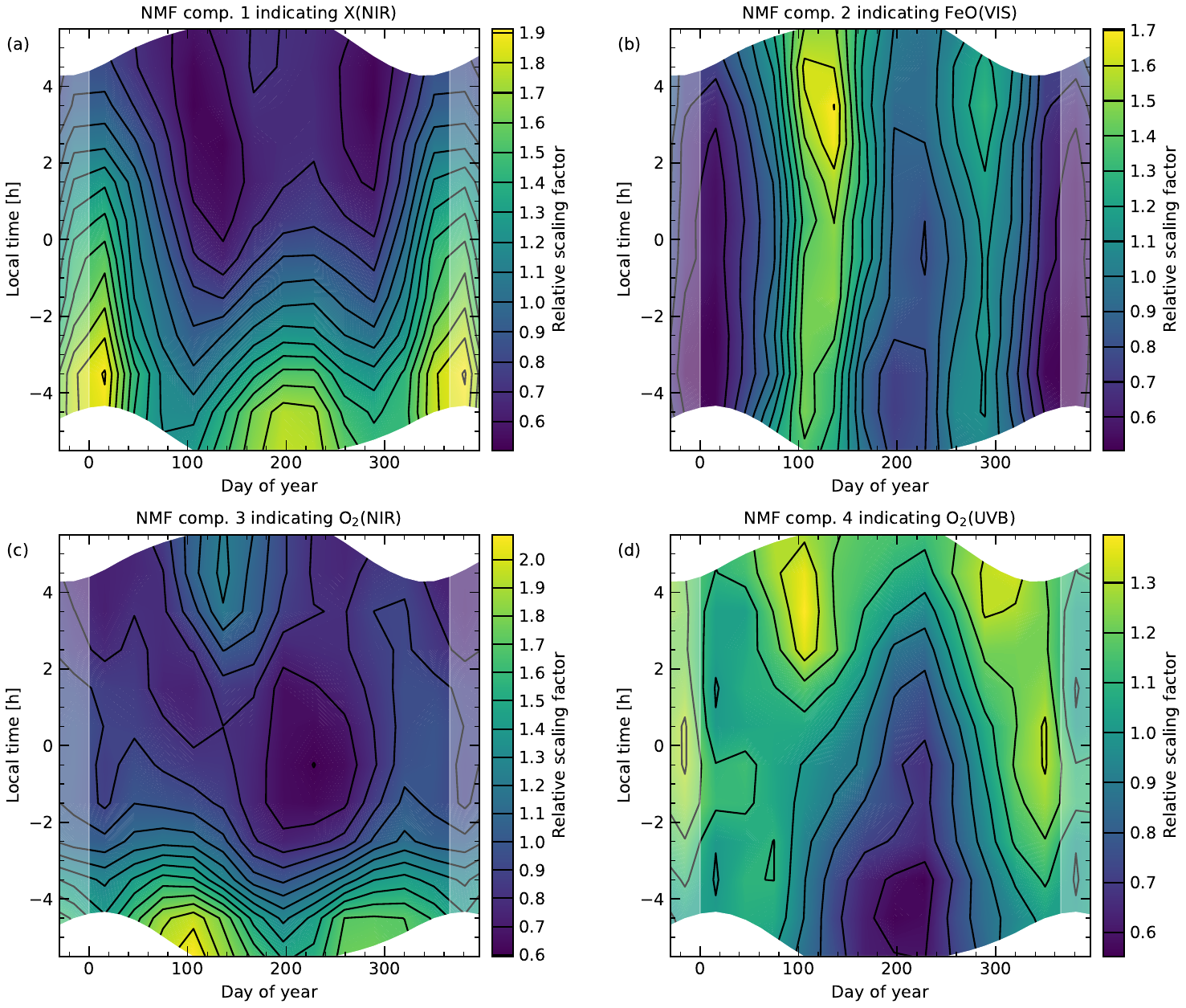}
\caption{Climatologies of the scaling factors of the four continuum
  components \chem{X}(NIR) (a), \chem{FeO}(VIS) (b), \chem{O_2}(NIR) (c), and
  \chem{O_2}(UVB) (d) from non-negative matrix factorisation shown in
  Fig.~\ref{fig:contcomp}a. Consistent with Fig.~\ref{fig:featclim_I}, the
  climatologies are also based on a sample of 7,971 30\,\unit{min} bins and a
  minimum subsample size of 200.}
\label{fig:compclim_I}
\end{figure*}

The two discussed features only cover a small part of the corresponding
NMF-related component spectra. In the studied wavelength ranges (see
Fig.~\ref{fig:contcomp}), \chem{FeO}(VIS) and \chem{X}(NIR) indicate mean
intensities of about 2.5 and 9.9\,\unit{kR} (explaining about 18 and 69\% of
the mean spectrum), i.e. the features at 595 and 1,510\,\unit{nm} have a
contribution of about 1.1 and 13.8\%. These percentages further decrease if
the radiance in the spectral gaps is roughly approximated by a simple linear
interpolation, which results in about 2.9 and 11.8\,\unit{kR} for the two
components. In particular, the \chem{X}(NIR) intensity could further increase
as the flux is still relatively high at the upper wavelength limit of
1,780\,\unit{nm}. Apart from the limited wavelength coverage, these values are
affected by uncertainties in the separation of the component spectra from
other contributions. Nevertheless, the basic structure of the \chem{FeO}(VIS)
and \chem{X}(NIR) components appears to be realistic. The 2D climatology of
the 595\,\unit{nm} feature is well correlated with the one of the underlying
continuum ($r = +0.961$). Moreover, the integrated flux between minima at 679
and 927\,\unit{nm} (Fig.~\ref{fig:meanfeat}) shows a high $r$ of $+0.974$. For
the 1,510\,\unit{nm} feature, the $r$ values are even above $+0.99$ if they
are calculated for the continuum below the feature or the secondary peak at
1,620\,\unit{nm} measured between 1,596 and 1,662\,\unit{nm}. Even in the
continuum below the \mbox{\chem{O_2}(a-X)(1-0)} band at about
1,080\,\unit{nm}, $r$ is still quite high with $+0.966$. Although partly
forced by the wavelength weighting of our NMF procedure, an $r$ of $+1.000$ is
remarkable for the correlation of the \chem{X}(NIR) component
(Fig.~\ref{fig:compclim_I}a) and the 1,510\,\unit{nm} peak
(Fig.~\ref{fig:featclim_I}b). The correlation of the \chem{FeO}(VIS) component
and the 595\,\unit{nm} peak is weaker ($+0.926$). The main difference in the
2D climatologies is a lower intensity for the NMF component in the evening
compared to the morning (Fig.~\ref{fig:compclim_I}b). This might be caused by
the decreasing nocturnal trend in the stronger \chem{X}(NIR) component, which
partly overlaps with \chem{FeO}(VIS). Thus, the NMF obviously led to more
different climatologies than the direct feature measurements showed.

The separation of the two \chem{O_2}-related components probably succeeded due
to a relatively weak SAO in the climatologies (panels (c) and (d) of
Fig.~\ref{fig:compclim_I}). The climatological patterns are more reminiscent
of the case for \chem{O} \citep{noll19} with tidal features that are also
visible in \chem{OH} intensity climatologies \citep{noll23}. This similarity
is reasonable as the nocturnal production process of these bands is probably
related to \chem{O} recombination \citep[e.g.,][]{slanger03} as well as
collisions of \chem{O_2} with excited oxygen atoms in the case of the near-IR
emissions \citep{kalogerakis19b}. Nevertheless, the correlation coefficient
for \chem{O_2}(UVB) and \chem{O_2}(NIR) is $-0.22$. The largest discrepancy is
present in the evening, when the intensity of \chem{O_2}(NIR) steeply
decreases due to the decay of the \chem{O_2}(a$^1\Delta_{\mathrm{g}}$)
population produced by \chem{O_3} photolysis at daytime
\citep[e.g.,][]{noll16}, whereas the intensity of \chem{O_2}(UVB) that is
related to electronic states without such a pathway is relatively low.
Interestingly, the excess \chem{O_2}(a$^1\Delta_{\mathrm{g}}$) population seems
to show an SAO which is consistent with a dependence on the \chem{O_3} density
as in the case of \chem{FeO}(VIS). The decrease of the intensity with
increasing LT might complicate the dynamical separation of \chem{O_2}(NIR) and
\chem{X}(NIR), which could contribute to the uncertainties around the
\mbox{\chem{O_2}(a-X)(0-0)} band at 1,270\,\unit{nm} in
Fig.~\ref{fig:contcomp}.

\subsection{Solar cycle effect}
\label{sec:SCE}

\begin{figure*}[t]
\includegraphics[width=17cm]{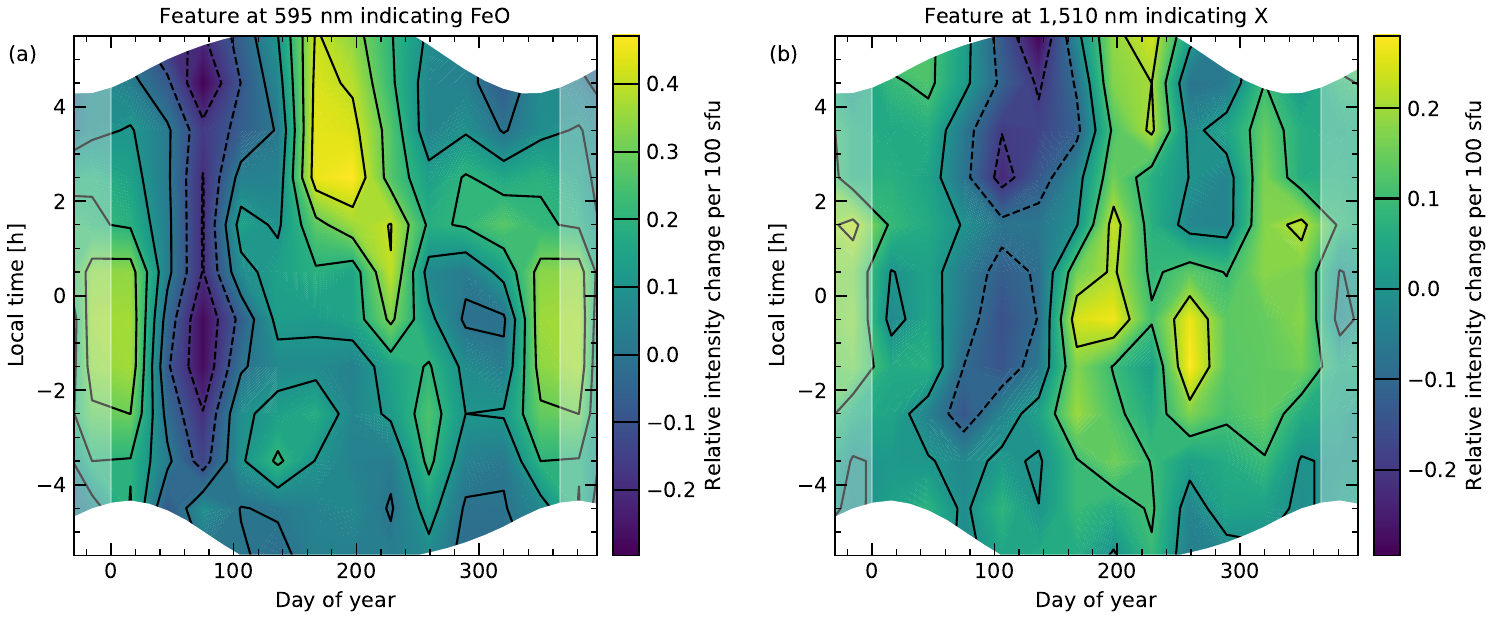}
\caption{Climatologies of the solar cycle effect for the continuum features at
  (a) 595\,\unit{nm} and (b) 1,510\,\unit{nm}. For each grid point (see
  caption of Fig.~\ref{fig:featclim_I}), the given value indicates the change
  of the intensity relative to the corresponding mean for an increase of the
  solar radio flux averaged for 27 days by 100\,\unit{sfu}. The climatologies
  were calculated for (a) 12,785 and (b) 17,482 30\,\unit{min} bins and the
  minimum sample size for each grid point was 400 (cf.
  Fig.~\ref{fig:featclim_I}).}
\label{fig:featclim_SCE}
\end{figure*}

As the \mbox{X-shooter} data set covers 10 years between October 2009 and
September 2019, the resulting continuum features can also be investigated with
respect to the solar cycle. As already discussed in Sect.~\ref{sec:intclim},
we also calculated 2D climatologies for the SCE. With respect to the features
at 595 and 1,510\,\unit{nm}, it turned out that the structures in these
climatologies are relatively uncertain. Based on the largest analysed sample
for the \chem{FeO} main peak with 12,785 bins, Fig.~\ref{fig:featclim_SCE}a
indicates the largest positive SCE values around the austral summer solstice
and in the austral winter. The lowest (and possibly negative) values appear to
be present around March. Figure~\ref{fig:featclim_SCE}b for the sample with
17,482 bins of the 1,510\,\unit{nm} feature shows possible maxima in July and
November and a minimum in austral autumn, which could possibly be negative.
Despite a low correlation coefficient of $+0.31$, the SCE climatologies of
both features appear to be more similar than in the case of the intensity
climatologies displayed in Fig.~\ref{fig:featclim_I}. 

The resulting effective SCEs derived from the averaging of the 595 and
1,510\,\unit{nm} climatologies are $+10.7$ and $+4.2$ \% per 100\,\unit{sfu},
respectively. If these percentages are directly derived from the intensities
of the 30\,\unit{min} bins, the results are $+8.1 \pm 1.5$ and
$+7.5 \pm 1.4$ \% per 100\,\unit{sfu}. For the individual measurements, we
obtain $+4.0 \pm 1.2$ and $+6.7 \pm 0.9$ \% per 100\,\unit{sfu}. The
differences between these results show the uncertainties related to the sample
size and the climatological weighting of the data points. In any case, the
SCEs for both continuum features are relatively small, which may explain the
relatively high uncertainties in the discussed climatological patterns. In
contrast, the \chem{O_2}-related features in the continuum show large effects
of about $+40$ \% per 100\,\unit{sfu} (e.g., using the ranges 335 to
388\,\unit{nm} and 1,254 to 1,297\,\unit{nm} for the NMF-related sample). For
\chem{OH}, the \mbox{X-shooter} data set indicates line-specific effective
SCEs between $+8$ and $+23$ \% per 100\,\unit{sfu} \citep{noll23}. On the
other hand, chemiluminescent 770\,\unit{nm} potassium (\chem{K}) emission
measured between April 2000 and March 2015 at Cerro Paranal in spectra of the
Ultraviolet and Visual Echelle Spectrograph \citep[UVES;][]{dekker00} resulted
in a negative effect of $-7.4 \pm 1.3$ \% per 100\,\unit{sfu}
\citep{noll19}. This seems to be related to an even more negative SCE
for the \chem{K} column density, as shown by WACCM simulations for the long
period from 1955 to 2005 \citep{dawkins16}. For the latitude range from 0 to
30$^{\circ}$\,S, about $-14.4$ \% per 100\,\unit{sfu} are given. The same study
also provides $-4.7$ \% per 100\,\unit{sfu} for the \chem{Fe} column density.
Considering that \chem{Fe} and \chem{K} react with \chem{O_3} to form monoxides
that are directly (\chem{FeO}) or indirectly (\chem{KO} with subsequent
reaction with \chem{O}) the basis for the chemiluminescence, the difference in
the SCEs for the column density of about 10\% would support a slightly
positive value for \chem{FeO} nightglow, which would be consistent with our
measurements (see also Sect.~\ref{sec:ressim}).

\subsection{Effective emission heights}
\label{sec:heights}

Using the \mbox{X-shooter} NIR-arm data set, \citet{noll22} investigated eight
nights in 2017 and seven nights in 2019 with respect to the signatures of
passing quasi-two-day waves (Q2DWs) in the intensities of \chem{OH} emission
lines. Q2DWs are only present for a few weeks in austral summer but constitute
the strongest wave phenomenon at low southern latitudes
\citep{ern13,gu19,tunbridge11}. The particularly strong wave between 26
January and 3 February 2017 was used to estimate the effective emission
heights of the selected 298 \chem{OH} lines based on fits of wave phases for a
most likely period of 44\,\unit{h}. Apart from the line intensities from the
\mbox{X-shooter} data, the study also used \chem{OH} emission profiles from
TIMED/SABER \citep{russell99} for the derivation of the required phase--height
relation. 

In order to better understand the emission features at 595 and
1,510\,\unit{nm}, which are obviously representative of a large fraction of
the nightglow continuum (Sect.~\ref{sec:intclim}), we also attempted to derive
wave phases and the related emission heights for these two features. For a
good time coverage during the crucial eight nights, we had to further relax
the selection criteria described in Sect.~\ref{sec:intclim}. Like the
investigated \chem{OH} lines, the 1,510\,\unit{nm} feature is covered by the
NIR arm. With a minimum exposure time of 1\,\unit{min} and only intensities
between 200 and 4,800\,\unit{R}, 265 of 388 observations remained in the
sample. We manually checked every spectrum and rejected 13 additional spectra
with suspicious astronomical targets, i.e. the final sample comprises
252 intensity measurements. Consistent with \citet{noll22}, the intensities of
30\,\unit{min} bins were calculated. If the default lower limit of
10\,\unit{min} for the bin filling is used, the resulting sample comprises 82
bins, which is lower than the maximum of 88 for the \chem{OH}-related sample.
However, the sample size can easily be increased to 92 bins if the bin filling
threshold is set slightly lower to 8\,\unit{min}. Then, only three bins of the
original \chem{OH}-related sample are lost, all of them present in the
evening. As the 595\,\unit{nm} feature is distinctly weaker and the smaller
sample of VIS-arm spectra has to be used, the final sample just comprises 125
spectra. The selection criteria include a minimum exposure time of
3\,\unit{min}, the requirement of positive values for the feature and the
underlying continuum (for the latter also an upper limit of
18\,\unit{R\,nm^{-1}}), and the rejection of additional spectra contaminated
by the astronomical targets (identification by visual inspection). Then, the
binning results in 63 bins if a minimum filling of 8\,\unit{min} is also
required (otherwise only 57 bins). 

The resulting bin-related intensities normalised by the sample mean were
fitted as described in \citet{noll22}. The fit formula
\begin{equation}\label{eq:Q2DWfit}
f(t, t_\mathrm{LT}) = 
c(t_\mathrm{LT}) \left(a(t_\mathrm{LT}) \cos\left(2\pi 
\left(\frac{t}{T}\,-\,\phi\right)\right) + 1\right),
\end{equation}
contains a cosine with the time $t$ relative to the period $T$ minus a
reference phase $\phi$ for 30 January 2017 12:00 LT. The cosine is multiplied
by an amplitude $c{\cdot}a$ and a constant $c$ (which can also be considered
as a scaling factor for the mean) is added to this term. As $T$ was set to
44\,\unit{h}, i.e. the optimum derived from the \chem{OH} data analysed by
\citet{noll22}, the final fitting parameters were $\phi$, $c$, and, $a$, the
latter being the amplitude of the cosine. As the \chem{OH} time series showed
a strong dependence of the amplitude $c{\cdot}a$ on local time, LT intervals
with a length of 1\,\unit{h} centred on $t_\mathrm{LT}$ were fitted separately.
First, this was done for the derivation of the optimum phase, which represents
the average for the selected LT hours weighted by the inverse of the phase
uncertainty. In a second step, the phase $\phi$ was fixed and the LT-dependent
parameters $c$ and $a$ were fitted.

\begin{figure}[t]
\includegraphics[width=8.3cm]{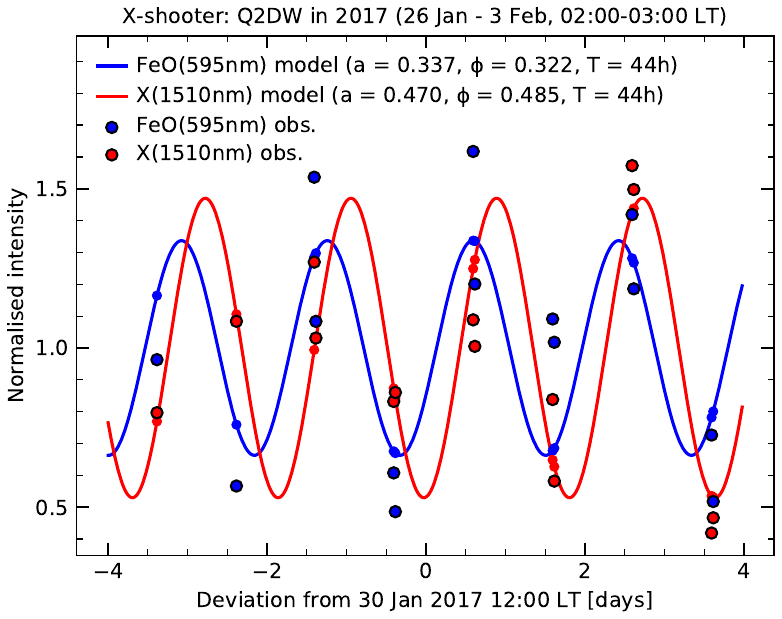}
\caption{Relative intensities of the features at 595\,\unit{nm} (blue) and
  1,510\,\unit{nm} (red) for the 14 30\,\unit{min} bins in the interval
  between 02:00 and 03:00 LT between 26 January and 3 February 2017 and the
  related fit of a cosine with a period $T$ of 44\,\unit{h} (solid curves with
  small dots for the effective times of the bins). Measurements and models are
  given relative to the fitted scaling factor $c$. The remaining fit
  parameters $a$ (amplitude) and $\phi$ (phase) are provided in the plot.}
\label{fig:Q2DWfit}
\end{figure}

For the derivation of the phase, only reliable LT intervals with a good time
coverage and small phase uncertainties should be used. \citet{noll22} selected
four to five LT hours depending on the line. The evening data were always
rejected because of low wave amplitudes and a relatively small number of bins.
For the 1,510\,\unit{nm} feature, the extended sample with 92 bins includes
the same 39 bins between 01:00 and 04:00 LT that were also considered for the
\chem{OH}-related fits. In the case of the 595\,\unit{nm} feature, the sample
with 63 entries shows 38 bins in this range. Alternatively, the fits can be
restricted to the interval between 02:00 and 03:00 LT, where all samples
include the same (and maximum number of) 14 bins. For the latter case,
Fig.~\ref{fig:Q2DWfit} shows the measured intensities and the resulting fits
divided by $c$ for both continuum features. This normalisation allows one to
directly read the amplitude $a$ from the plotted wave fit. The amplitude is
higher for the 1,510\,\unit{nm} feature (0.47 vs. 0.34). This feature also
shows a later phase (0.485 vs. 0.322 relative to $T$). Concerning the fit
quality, it is clearly visible that the deviations for the strong
1,510\,\unit{nm} feature are distinctly smaller than in the case of the
595\,\unit{nm} feature. The root mean square results in 0.14 compared to 0.23.
Nevertheless, the phase for the \chem{FeO} main peak does not seem to be less
robust since the standard deviation for the independent fits of the three
intervals between 01:00 and 04:00 LT indicates 0.033 compared to 0.042 for the
peak at 1,510\,\unit{nm}. The mean phase from these intervals is slightly
higher in both cases (0.489 and 0.339).

\begin{figure}[t]
\includegraphics[width=8.3cm]{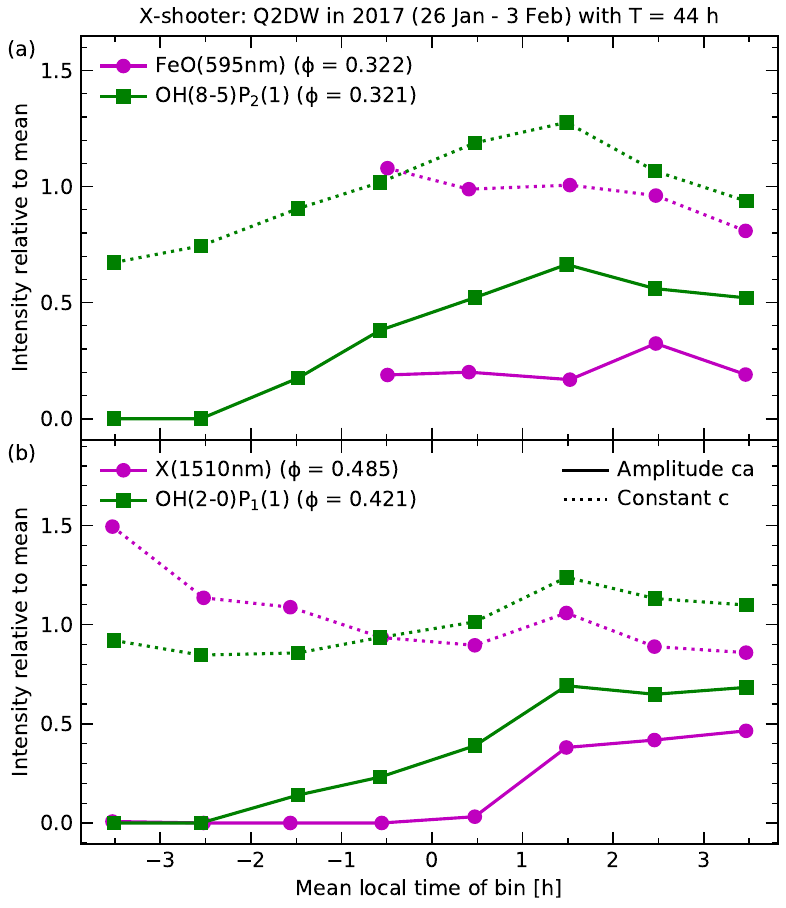}
\caption{Amplitudes $c{\cdot}a$ (magenta solid curves with circles) and
  scaling constants $c$ (magenta dotted curves with circles) relative to the
  sample mean as a function of local time (step size of 1\,\unit{h}; only
  intervals with sufficient data) for cosine fits of the Q2DW in 2017 with a
  period of 44\,\unit{h} based on intensity data of the (a) 595\,\unit{nm}
  (sample of 63 bins) and (b) 1,510\,\unit{nm} (sample of 92 bins) features.
  The given phases are only based on fits of the interval between 02:00 and
  03:00 LT. In addition, the corresponding curves for \chem{OH} emission lines
  (green curves with squares) with similar phases $\phi$ at 30 January 12:00
  LT from the same fitting procedure are shown \citep[cf.][]{noll22}.}
\label{fig:Q2DWampl}
\end{figure}

For both continuum features, Fig.~\ref{fig:Q2DWampl} shows the LT-dependent
amplitudes $c{\cdot}a$ and scaling constants $c$ for an optimum phase $\phi$
that is only based on the interval centred on 02:30 LT. For 595\,\unit{nm} in
(a), there were only sufficient data (at least seven bins) for a fit in the LT
range between 23:00 and 04:00 LT. Hence, the situation in the evening remains
unclear. For the covered time range, the amplitude relative to the mean is
about 0.2 with a peak of 0.32 for 02:00 to 03:00 LT, i.e. the decisive
interval for the phase derivation. The constant $c$ is around 1, which
indicates that there was not a clear trend of the mean with local time. We
compare these curves with those of an \chem{OH} line with almost the same
phase for the 02:30 LT interval. Note that the given $\phi$ of 0.321 is
slightly lower than the value of 0.328 in the data release of
\citet{noll22ds}, which is based on several LT hours. The data for
\mbox{\chem{OH}(8-5)P$_2$(1)} indicate a significantly larger maximum
amplitude $c{\cdot}a$ of 0.71 (between 01:00 and 02:00 LT). Even if the
different scaling factors are considered and only the cosine amplitudes $a$
are compared (0.54 vs. 0.34 between 02:00 and 03:00 LT), the impact of the
Q2DW on \chem{OH} lines appears to be stronger, which probably reveals
dynamical differences if \chem{O_3} reactions with \chem{Fe} and \chem{H} are
compared. The remarkable decrease of the wave amplitude towards the beginning
of the night is not covered by the data for the 595\,\unit{nm} feature. Hence,
a clear difference in the nocturnal trends in the joint LT range is not
obvious.

The 1,510\,\unit{nm} data in Fig.~\ref{fig:Q2DWampl}b show the entire
\chem{OH}-related night interval. The scaling factor $c$ indicates a clear
decrease in the course of the night, which implies that the average nocturnal
behaviour as shown in the climatology in Fig.~\ref{fig:featclim_I}b is also
relevant for the eight nights affected by the strong Q2DW. On the other hand,
there is no detection of this wave except for the last three intervals between
01:00 and 04:00 LT, which explains why only this range was useful for phase
fits. There $c{\cdot}a$ is clearly larger than for the \chem{FeO} main peak
(like $a$ in Fig.~\ref{fig:Q2DWfit}). We compare the 1,510\,\unit{nm} feature
with \mbox{\chem{OH}(2-0)P$_1$(1)}, which shows the highest phase for the
studied \chem{OH} lines. The nocturnal trend in $c{\cdot}a$ of this line seems
to be roughly consistent, although the increase in the middle of the night
starts earlier and is slower. These differences might be related to the
remaining $\Delta\phi$ of 0.064. The discrepancies between the absolute
$c{\cdot}a$ of continuum feature and \chem{OH} line appear to be mainly
related to the different nocturnal mean behaviour. The deviation of the
1,510\,\unit{nm} feature with respect to the cosine amplitude $a$ is
relatively small at the end of the night (0.54 vs. 0.62 for 03:00 to 04:00
LT), i.e. the responses to the passing Q2DW appear to be comparable. The
agreement in the nocturnal development of the amplitude also suggests that the
high phase value for the 1,510\,\unit{nm} feature is realistic. 

As described by \citet{noll22}, the \mbox{X-shooter}-based reference phases of
the Q2DW were converted into heights by using the linear phase--height
relation derived from altitude-dependent wave fits of 22 \chem{OH} emission
profiles in the SABER 2.1\,\unit{\mu{}m} channel \citep{russell99}, which were
taken around Cerro Paranal in the eight relevant nights at about 04:00 LT. The
regression for the height range from 80 to 97\,\unit{km} resulted in an
intercept of $3.027 \pm 0.049$ at 0\,\unit{km} and a vertical wavelength of
$31.74 \pm 0.56$\,\unit{km}. \citet{noll22} also applied a correction that
considers differences in the properties of the \mbox{X-shooter} and SABER
samples. Based on a comparison of phase fits for the vertically integrated
emission profiles for both \chem{OH}-related SABER channels and the phases
from the \mbox{X-shooter} data weighted for the transmission curves of these
channels, a general shift of the heights by $-0.43 \pm 0.13$\,\unit{km} was
performed. As the phases slightly change if only the LT interval between 02:00
and 03:00 is used for their derivation as described above, we recalculated
this shift and found about $-0.79$\,\unit{km}. As a result (which also
considers the systematic decrease in $\phi$), the mean effective emission
height of all 298 \chem{OH} lines was 0.10\,\unit{km} lower than given by
\citet{noll22}. If the mean of the phases from the three intervals between
01:00 and 04:00 LT is used instead, the resulting offset of $-0.45$\,\unit{km}
is very close to the original value and the mean height decreases just by
0.05\,\unit{km}. Hence, the change in the calculation of the optimum phase
does not appear to have a significant effect on the resulting emission
heights. Using the SABER-based phase--height relation and
\mbox{X-shooter}-based phases for the 02:30 LT interval with the estimated
corrections (i.e. $-0.79$\,\unit{km} and $+0.1$\,\unit{km} to be consistent
with the published \chem{OH}-related heights), we finally obtain altitudes of
85.2 and 80.0\,\unit{km} for the 595 and 1,510\,\unit{nm} features,
respectively. From the comparison of heights for \chem{OH} lines with the same
or similar ro-vibrational upper levels, \citet{noll22} found uncertainties of
several tenths of a kilometre. If we take the reported phase standard
deviations of about 0.04 as derived from the LT hours between 01:00 and 04:00
for the two continuum features as an indicator, then the uncertainties might
even be of the order of 1\,\unit{km}. Moreover, the use of the full time
series is important for the quality of the results. For example, the heights
would be unrealistically high (near 100\,\unit{km}) if the first night was
excluded from the fits. However, the difference between the values for both
features would only slightly change and rejecting the last night would only
have a minor effect. 

The given altitudes are representative of the effective height for the
strongest absolute variations related to the passing Q2DW. They differ from
the effective mean height of the emission. \citet{noll22} found that the
average centroid emission altitude for the two \chem{OH}-related SABER
channels at Cerro Paranal \citep{noll17} was about 4.07\,\unit{km} higher than
in the case of the corresponding variability-related heights. Significant
discrepancies are not surprising in this context since the steep decrease of
the \chem{O} number density in the lower parts of the \chem{OH} emission
profile \citep[e.g.,][]{smith10} lead to stronger relative intensity
variations towards lower heights. Nevertheless, the amount of the discrepancy
is quite high, which might be explained by the large amplitude of the Q2DW. It
is questionable whether the effective emission heights for the continuum
features need to be shifted by a similar value. However, as \chem{OH} and
\chem{FeO} are produced by reactions that involve \chem{O_3}, it is not
unlikely that the impact of the \chem{O} profile is similar for the
595\,\unit{nm} peak at least. Moreover, the variability-related height for
\chem{FeO} is well in the range between 81.8 and 89.7\,\unit{km} found for the
\chem{OH} lines by \citet{noll22}. Thus, also assuming a shift of
4.07\,\unit{km}, we would obtain a centroid altitude of 89.2\,\unit{km}. This
value appears to be close to other observations. For the \chem{FeO} orange
bands, \citet{evans10} measured with OSIRIS on Odin between 0 and
40$^{\circ}$\,S in April/May 2003 a centroid emission height slightly (up to
1\,\unit{km}) higher than the peak at about 87\,\unit{km}. Modelling of the
\chem{FeO} layer involving FeMOD \citep{gardner05} at 20$^{\circ}$\,N in March
2000 even resulted in a peak height of 89.5\,\unit{km} \citep{saran11}.

Without a good knowledge of the chemistry related to the 1,510\,\unit{nm}
feature, the difference between mean centroid and Q2DW-related effective
emission height is uncertain. If the \chem{OH}-based shift is applied, the
former would be about 84.1\,\unit{km}. This is possibly an upper limit and
indicates that the emission layer appears to be lower than the \chem{OH} and
\chem{FeO} layers. Previous simulations of the \chem{Fe}-related layers with
WACCM by \citet{feng13} showed that the densities of neutral molecular
reservoir species such as \chem{FeO_3}, \chem{FeOH}, and \chem{Fe(OH)_2} can
peak several kilometres lower than the \chem{FeO} density. Modelling also
suggests that the \chem{HO_2} density maximises near 80\,\unit{km}
\citep{makhlouf95}. Such altitudes were also obtained from nocturnal microwave
measurements of \chem{HO_2}, although the density peaks could also be several
kilometres higher in the second half of the night \citep{kreyling13,millan15}.
Overall, the discussed candidates for emitter \chem{X} appear to produce
emission at heights consistent with our measurements. With our optimised WACCM
simulations, we can discuss the height distributions more quantitatively (see
Sect.~\ref{sec:ressim}).

\section{Modelling}
\label{sec:modelling}

\subsection{Model set-up}
\label{sec:modelsetup}

For a better understanding of the \mbox{X-shooter}-based nightglow continuum
and its variability as discussed in Sect.~\ref{sec:results}, we performed
dedicated WACCM simulations. Community Earth System Model (CESM1, WACCM4)
simulations with metal chemistry have previously been used for combined
observational and modelling studies of chemiluminescent \chem{FeO}
\citep{unterguggenberger17} and \chem{K}(4$^2$P) emissions \citep{noll19}
above Cerro Paranal. Here, we carried out modelling simulations from the
updated version of CESM2 (WACCM6) with \chem{Na} and \chem{Fe} chemistry to
check the \chem{FeO}-related results and to explore potential candidates for
the new pseudo-continuum. CESM2 (WACCM6) is described by \citet{gettelman19}.
\chem{Na} and \chem{Fe} chemistry is updated based on \citet{plane15}. The
meteoric injection function (MIF) of \chem{Fe} is from \citet{carrillo16},
which is different to that used in \citet{feng13}. We divided the \chem{Fe}
MIF by 5 to match lidar observations \citep[e.g.,][]{daly20}. Here, we use the
specified dynamics version of WACCM6 nudged with NASA's Modern Era
Retrospective Analysis for Research and Application MERRA2 reanalysis data set
\citep{molod15} below 50\,\unit{km}. Then, the amount of relaxation is
linearly reduced. Above 60\,\unit{km}, the model is free running
\citep[cf.][]{marsh13}. The model has a resolution of 1.9$^{\circ}$ in latitude
and 2.5$^{\circ}$ in longitude and contains 88 vertical levels from the
surface to 140\,\unit{km}. The simulation covers the period from 1 Jan 2003 to
28 Dec 2014 (Universal Time). Monthly mean values of selected variables were
calculated to save disc space. The model output was also sampled every half an
hour for 24$^{\circ}$\,S and 70$^{\circ}$\,W near Cerro Paranal and
interpolated in the height range from 40 to 130\,\unit{km} with a step size of
1\,\unit{km}. The latter is used for the analysis in Sect.~\ref{sec:ressim}.

\begin{table*}[th!]
\caption{Emission from electronically excited \chem{Fe}-containing molecules}
\begin{tabular}{clll}
\tophline
Number$^\mathrm{a}$ & Reaction & Rate coefficient$^\mathrm{b}$ & Reference \\
& & (\unit{cm^3 molecule^{-1} s^{-1}}) & \\
\middlehline
R2 & $\mathrm{Fe} + \mathrm{O}_3 \rightarrow \mathrm{FeO}^{\ast} +
\mathrm{O}_2$ & $2.9 \times 10^{-10} e^{-174/T}$ & \citet{feng13} \\
R6 & $\mathrm{FeOH} + \mathrm{O}_3 \rightarrow \mathrm{OFeOH}^{\ast} +
\mathrm{O}_2$ & $7.3 \times 10^{-10} (-200/T)^{-1.65}$ & This work \\
R10 & $\mathrm{FeO} + \mathrm{O}_3 \rightarrow \mathrm{FeO}_2^{\ast} +
\mathrm{O}_2$ & $3.0 \times 10^{-10} e^{-177/T}$ & \citet{rollason00} \\
R11 & $\mathrm{OFeOH} + \mathrm{O} \rightarrow \mathrm{FeOH} + \mathrm{O}_2$ &
$6.0 \times 10^{-10} (-200/T)^{-1.68}$ & This work \\
R12 & $\mathrm{OFeOH} + \mathrm{FeOH} \rightarrow \mathrm{MSP}$$^\mathrm{c}$ &
$9.0 \times 10^{-10}$ & This work \\
R13 & $\mathrm{OFeOH} + \mathrm{OFeOH} \rightarrow \mathrm{MSP}$$^\mathrm{c}$ &
$9.0 \times 10^{-10}$ & This work \\
\bottomhline
\end{tabular}
\belowtable{
\begin{list}{}{}
\item[$^\mathrm{a}$] consistent with numbering in text
\item[$^\mathrm{b}$] temperature $T$ in Kelvin
\item[$^\mathrm{c}$] meteoric smoke particles
\end{list}
}
\label{tab:chemFe}
\end{table*}

\begin{table*}[th!]
\caption{Potential mechanisms for generating \chem{HO_2} emission}
\begin{tabular}{clll}
\tophline
Number$^\mathrm{a}$ & Reaction & Rate coefficient & Reference \\
& & (\unit{cm^3 molecule^{-1} s^{-1}}) & \\
\middlehline
R8 & $\mathrm{HO}_2 + \mathrm{O}_2(\mathrm{a}^1\Delta_{\mathrm{g}}) \rightarrow
\mathrm{HO}_2^{\ast} + \mathrm{O}_2$ & $1.0 \times 10^{-10}$ & This work \\
R9 & $\mathrm{H} + \mathrm{O}_2 + \mathrm{M} \rightarrow \mathrm{HO}_2^{\ast} +
\mathrm{M}$ & $k_0 (4.4 \times 10^{-32}, n = 1.3),$ & \citet{burkholder15} \\
& & $k_\infty (7.5 \times 10^{-11}, m = -0.2)$$^\mathrm{b}$ & \\
R14 & $\mathrm{H} + \mathrm{O}_3 \rightarrow \mathrm{HO}_2^{\ast} + \mathrm{O}$
& $7.0 \times 10^{-13}$ (upper limit) & \citet{howard80} \\
R15 & $\mathrm{HO}_2 + \mathrm{O} \rightarrow \mathrm{OH} +
\mathrm{O}_2(\mathrm{a}^1\Delta_{\mathrm{g}})$ &
$0.95 \times 2.7 \times 10^{-11} e^{222.5/T}$ & This work$^\mathrm{c}$ \\
R16 & $\mathrm{HO}_2 + \mathrm{O} \rightarrow \mathrm{OH} +
\mathrm{O}_2(\mathrm{X}^3\Sigma_{\mathrm{g}}^{-})$ &
$0.05 \times 2.7 \times 10^{-11} e^{222.5/T}$ & This work$^\mathrm{c}$ \\
\bottomhline
\end{tabular}
\belowtable{
\begin{list}{}{}
\item[$^\mathrm{a}$] consistent with numbering in text
\item[$^\mathrm{b}$] low- and high-pressure limits with exponents $n$ and $m$
  for $(T_\mathrm{ref}/T)$ with $T_\mathrm{ref} = 300$\,\unit{K} (three-body
  recombination)
\item[$^\mathrm{c}$] c.f. \citet{burkholder15}
\end{list}
}
\label{tab:chemHO2}
\end{table*}

Table~\ref{tab:chemFe} lists potential \chem{Fe}-related nightglow chemistry
that was explored. Reaction~\ref{eq:Fe+O3} generates \chem{FeO} emission
\citep{feng13}. According to \citet{helmer94}, it is exothermic by
$301 \pm 8$\,\unit{kJ\,mol^{-1}}, i.e. almost the entire wavelength range of
the \mbox{X-shooter} nightglow spectrum could be covered. Reaction~R10 that
produces \chem{FeO_2} indicates a similar exothermicity of
$311 \pm 48$\,\unit{kJ\,mol^{-1}} \citep{rollason00}. For the other reactions
in Table~\ref{tab:chemFe}, the reaction exothermicities were calculated using
the high accuracy complete basis set CBS-QB3 method \citep{frisch16}. The
production of \chem{OFeOH} via Reaction~\ref{eq:FeOH+O3} is sufficiently
exothermic by 339\,\unit{kJ\,mol^{-1}} if
\chem{O_2}($\mathrm{X}^3\Sigma_{\mathrm{g}}^-$) is produced, or
244\,\unit{kJ\,mol^{-1}} if \chem{O_2}($\mathrm{a}^1\Delta_{\mathrm{g}}$) is the
product (which is the spin-conserving channel). \chem{OFeOH} has an abundance
of low-lying electronic states (eight states below 2.5\,\unit{eV}) which could
be involved in emission from the visual to the near-IR. Reaction~R11 is
exothermic by 60\,\unit{kJ\,mol^{-1}}. Both reactions should not have
significant barriers and hence calculated capture rate coefficients
\citep{georgievskii05} are assigned to these reactions. Reactions~R12 and R13
represent polymerisation of \chem{OFeOH} and \chem{FeOH} to make meteoric
smoke particles, which are treated as a permanent sink.

The first three reactions in Table~\ref{tab:chemHO2} constitute potential
mechanisms for the generation of excited \chem{HO_2}, an attractive candidate
for the source of the \chem{X}(NIR) component of the nightglow continuum
measured by \mbox{X-shooter}. We list the already mentioned
Reaction~\ref{eq:HO2+O2a} that is most relevant for the production of
chemiluminescent emission in the laboratory and Reaction~\ref{eq:H+O2+M} that
also produces chemiluminescence and is the main production process of
\chem{HO_2} (Sect.~\ref{sec:contdecomp}). In principle, the direct radiative
recombination of \chem{H} and \chem{O_2} could also contribute. However, this
mechanism is very unlikely to compete with the termolecular
Reaction~\ref{eq:H+O2+M} at the relatively high pressures of the mesopause
region. It would need a probability of the order of $10^{-8}$ to make a small
contribution at least. As confirmed by tests, the intensity variations from
this reaction are very similar to those of the termolecular case. Hence,
an estimate of the relevance would be very challenging. We neglect this
possible minor channel in the following. Lastly, the reaction between \chem{H}
and \chem{O_3} (Reaction~R14) is sufficiently exothermic to produce excited
\chem{HO_2}. However, the channel producing \chem{HO_2} + \chem{O} is known
from experiment to be minor (3\%) compared with the channel producing
\chem{OH} + \chem{O_2} \citep{howard80}.

Reactions~R15 and R16 lead to the destruction of \chem{HO_2} by collisions
with \chem{O}. The difference between both reactions is the electronic state
of the resulting \chem{O_2} molecule. We included this distinction as we
consider Reaction~R15 as the source of the nighttime production of
$\mathrm{a}^1\Delta_{\mathrm{g}}$. With a branching ratio of 95\%, we explore
the maximum possible contribution of this pathway. For a better understanding
of the impact of this percentage, we also performed a simulation with a
relative $\mathrm{a}^1\Delta_{\mathrm{g}}$ yield of 40\%. The main effect is
the decrease of the nocturnal \chem{HO_2} emission due to
Reaction~\ref{eq:HO2+O2a} around midnight and later in agreement with the
reduced percentage. At the beginning of the night, the impact is smaller as
$\mathrm{a}^1\Delta_{\mathrm{g}}$ mainly originates from the daytime \chem{O_3}
photolysis, which is considered in WACCM. The decay of this population shows
a time constant of the order of 1\,hour in the mesopause region
\citep{noll16}. There are other possible mechanisms that could contribute to
the nighttime $\mathrm{a}^1\Delta_{\mathrm{g}}$ population. However, there is a
remarkable lack of knowledge with respect to the efficiency of these
reactions. The `classical' pathway via \chem{O} recombination and subsequent
collisions \citep[e.g.,][]{barth61,slanger03} that is important for the
production of the $\mathrm{b}^1\Sigma_{\mathrm{g}}^{+}$ and higher electronic
states does not appear to lead to a sufficient $\mathrm{a}^1\Delta_{\mathrm{g}}$
population, including heights around the peak of the
\mbox{\chem{O_2}(b-X)(0-0)} band at 762\,\unit{nm} near 94\,\unit{km}
\citep[e.g.,][]{yee97}. \mbox{\chem{O_2}(a-X)(0-0)} at 1,270\,\unit{nm} shows
a mean centroid emission height of about 89\,\unit{km} at Cerro Paranal
\citep{noll16}. This altitude is similar to the centroid emission heights of
\chem{OH} lines, particularly those with relatively high vibrational
excitation \citep{noll22}. Therefore, reactions between \chem{OH} and \chem{O}
might lead to the generation of excited \chem{O_2} molecules in the altitude
range of the \chem{OH} emission. Another source of
$\mathrm{a}^1\Delta_{\mathrm{g}}$ at even lower heights could be
Reaction~\ref{eq:H+O2+M} if \chem{M} is \chem{O_2}. However, the efficiency is
quite uncertain. Hence, we focus on Reaction~16, which also provides
$\mathrm{a}^1\Delta_{\mathrm{g}}$ at heights relevant for \chem{HO_2}. We
evaluate this choice in Sect.~\ref{sec:ressim}.

To be consistent with the analysis of the nocturnal/seasonal variations in the
\mbox{X-shooter}-based measurements (Sect.~\ref{sec:intclim}), we derived
model climatologies for the vertically integrated volume emission rates
between 40 and 130\,\unit{km} in a similar way. We only considered nighttime
data with solar zenith angles larger than 100$^{\circ}$, which reduced the
sample from 210,240 to 92,064 time steps for 12 years. The time resolution of
30\,\unit{min} is consistent with the binning of the \mbox{X-shooter} data. As
the resulting sample is still much larger than those used for the
climatologies of the binned measured data, it is was not necessary to partly
decrease the resolution to achieve minimum subsample sizes of 200 or 400 for
each relevant grid point (Sect.~\ref{sec:intclim}). Consistent with the
\mbox{X-shooter}-related results, the intensity climatologies are given for a
fixed solar radio flux averaged for 27 days of 100\,\unit{sfu}. Although the
covered time interval from 2003 to 2014 only partly overlaps with the
\mbox{X-shooter} sample (Oct 2009 to Sep 2019), the mean solar radio flux of
all nighttime climatological grid points was also very close to this reference
value (97 to 107\,\unit{sfu} with a mean of 100\,\unit{sfu}), which caused
only very small corrections. We also compared the effective solar cycle
effects for the nighttime climatologies of the measured and modelled
emissions. The shift of the time interval certainly contributes to the
systematic uncertainties but does not appear to significantly increase them.
Based on the results for the different calculation methods discussed in
Sect.~\ref{sec:SCE}, we expect total uncertainties of the order of a few per
cent per 100\,\unit{sfu}. Apart from intensity climatologies, we also derived
climatologies for the centroid heights of the emissions, i.e.
emission-weighted heights for the whole range from 40 to 130\,\unit{km}. The
typical nighttime emission profiles were mostly well localised in the
mesopause region. Finally, we calculated climatologies for the number
densities of different relevant atmospheric species at specific heights.

\subsection{Results from simulations}
\label{sec:ressim}

\begin{figure*}[t]
\includegraphics[width=17cm]{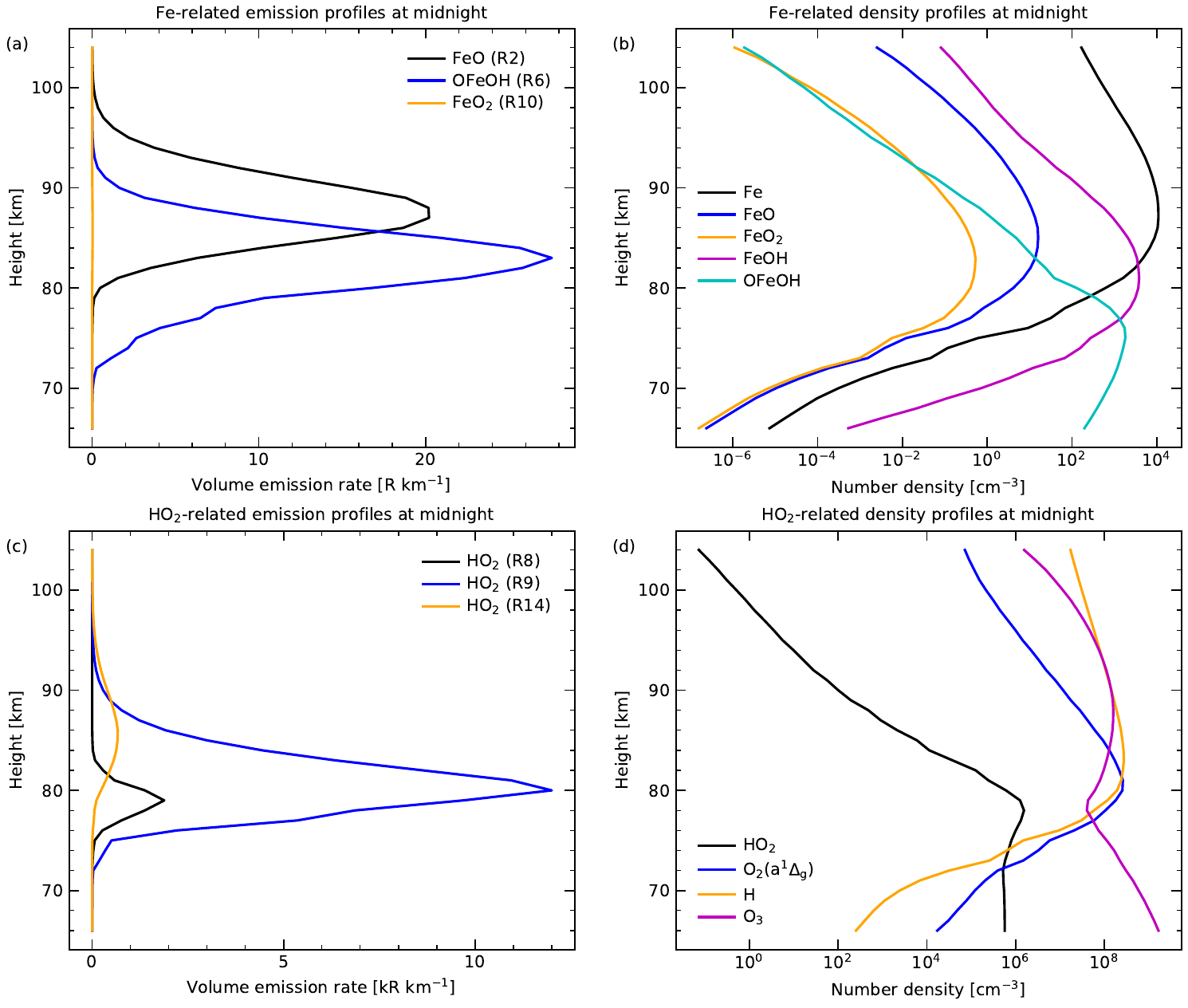}
\caption{WACCM-related mean profiles of (a) volume emission rates of excited
  \chem{Fe}-containing molecules (see Table~\ref{tab:chemFe}), (b) number
  densities of \chem{Fe}-containing molecules, (c) volume emission rates of
  excited \chem{HO_2} produced by different processes (see
  Table~\ref{tab:chemHO2}), and (d) number densities of species relevant for
  the production of excited \chem{HO_2} for local midnight (i.e. 8,760
  profiles with local times of 23:48 and 00:18) at 24$^{\circ}$\,S and
  70$^{\circ}$\,W.}
\label{fig:simprof}
\end{figure*}

\begin{figure*}[t]
\includegraphics[width=17cm]{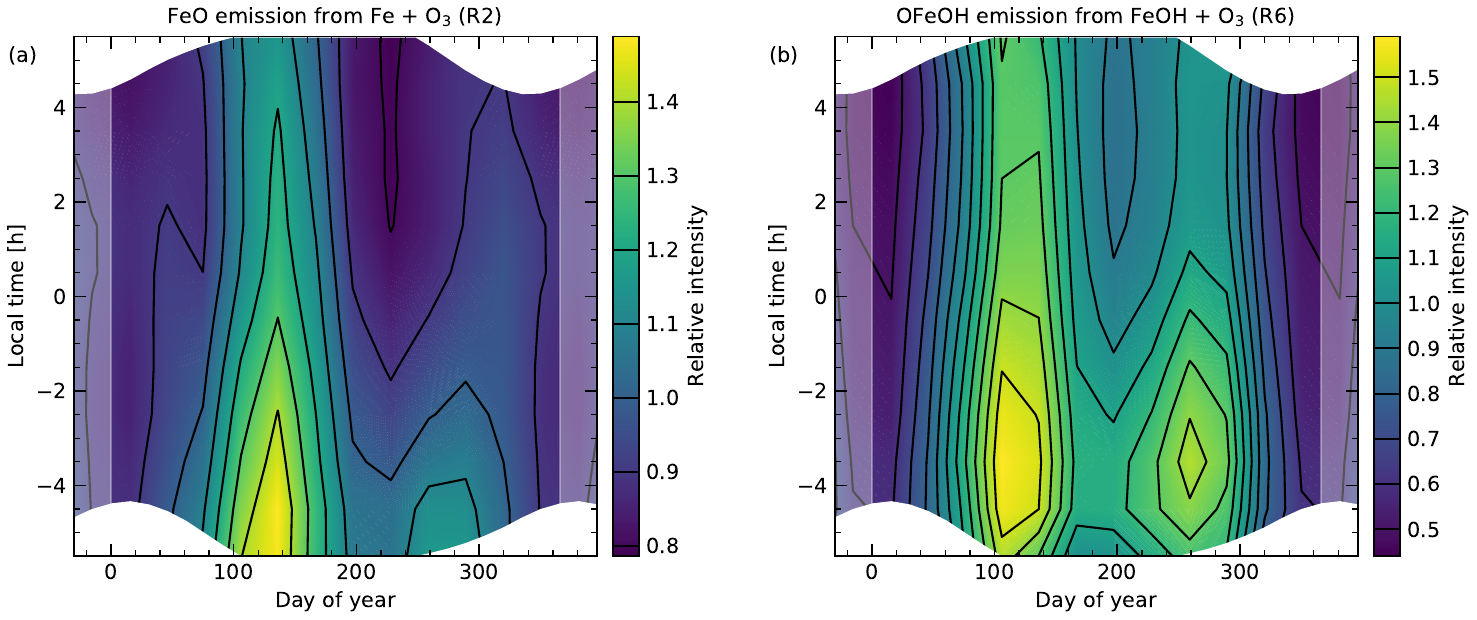}
\caption{Climatologies of intensity relative to the mean for the
  vertically-integrated WACCM-simulated emission of (a) \chem{FeO} and (b)
  \chem{OFeOH} (see Table~\ref{tab:chemFe}) at 24$^{\circ}$\,S and
  70$^{\circ}$\,W. The climatologies were derived in a similar way as those in
  Fig.~\ref{fig:featclim_I}. With 92,064 selected nighttime data points, the
  subsample size for the relevant grid points was well above the limit of 200
  (and even 400) used for the \mbox{X-shooter} data.}
\label{fig:simclim_Fe_I}
\end{figure*}

\begin{figure*}[t]
\includegraphics[width=17cm]{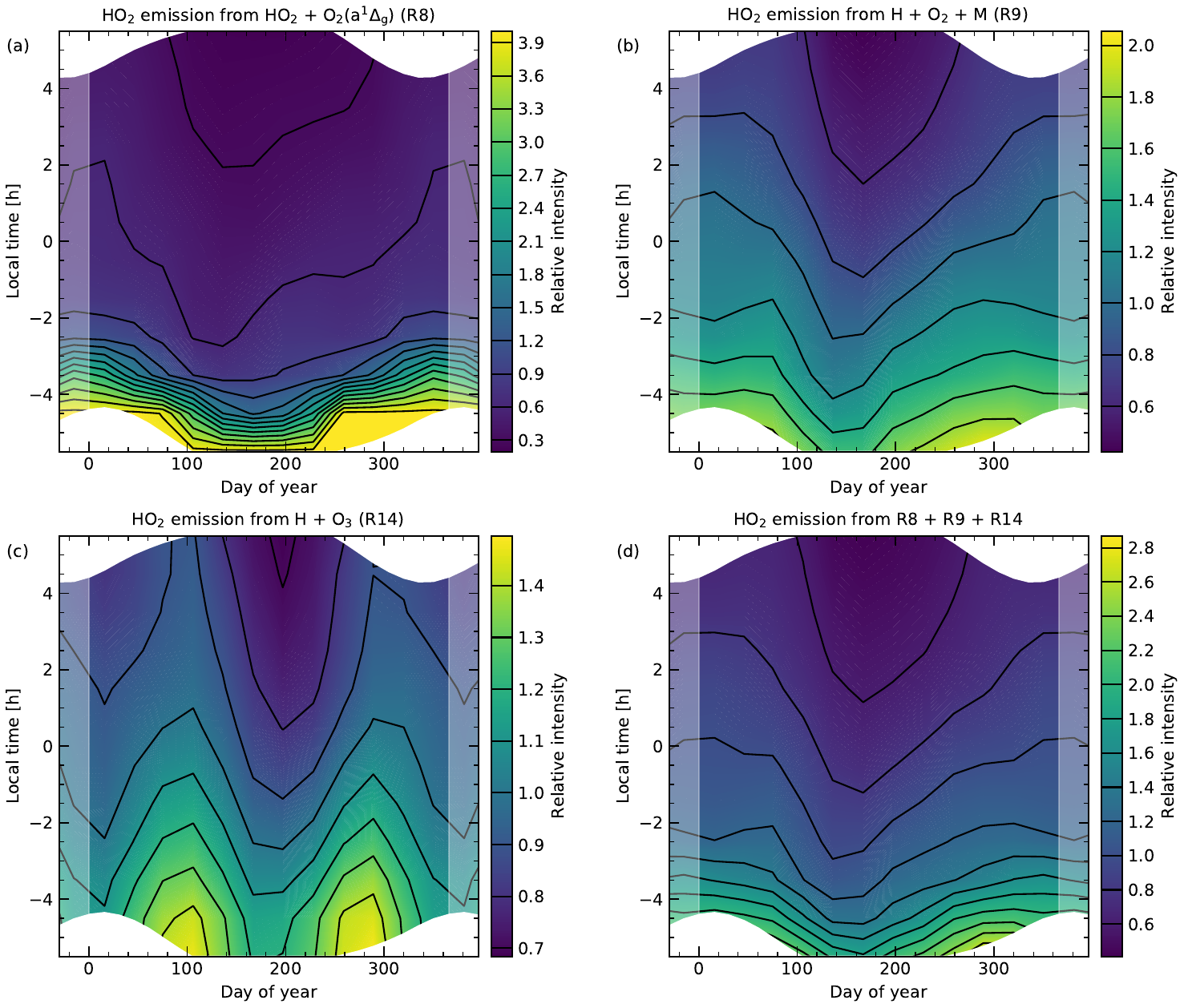}
\caption{Climatologies of intensity relative to the mean for the
  vertically-integrated WACCM-simulated emission of excited \chem{HO_2}
  produced by three different processes (see Table~\ref{tab:chemHO2}) and the
  sum of them at 24$^{\circ}$\,S and 70$^{\circ}$\,W. Sample and calculation of
  the climatologies were the same as for Fig.~\ref{fig:simclim_Fe_I}. As the
  climatology in (a) shows a very large dynamical range, the colour scale was
  cut at a relative intensity of 4, which decreased the visualised range by
  more than a factor of 2.}
\label{fig:simclim_HO2_I}
\end{figure*}

\begin{table*}[t]
\caption{Comparison of nightglow continuum emissions from \mbox{X-shooter}
 spectra and WACCM simulations of potential emission processes}
\begin{tabular}{lccccc}
\tophline
Emission$^\mathrm{a}$ & $\langle I \rangle$$^\mathrm{b}$ &
$r_\mathrm{f06a}$$^\mathrm{c}$ & $r_\mathrm{f15a}$$^\mathrm{d}$ &
$\langle h_\mathrm{cen} \rangle$$^\mathrm{e}$ &
$\langle \mathrm{SCE} \rangle$$^\mathrm{f}$ \\
& (\unit{kR}) & & & (\unit{km}) & ($10^{-2}$\,\unit{sfu^{-1}}) \\
\middlehline
595\,\unit{nm} (f06a) & 0.027 & $+1.000$ & $-0.305$ & 85.2--89.2 & $+0.107$ \\
\chem{FeO}(VIS) & 2.90 & $+0.926$ & $-0.600$ & & \\
\chem{FeO} (R2) & 0.170 & $+0.807$ & $+0.116$ & 87.9 & $+0.158$ \\
\chem{OFeOH} (R6) & 0.220 & $+0.867$ & $-0.190$ & 82.3 & $+0.053$ \\
\chem{FeO_2} (R10) & 0.0002 & $+0.744$ & $+0.262$ & 85.8 & $+0.102$ \\
\middlehline
1,510\,\unit{nm} (f15a) & 1.37 & $-0.305$ & $+1.000$ & 80.0--84.1 & $+0.042$ \\
\chem{X}(NIR) & 11.81 & $-0.320$ & $+1.000$ & & \\
\chem{HO_2} (R8) & 12.74 & $+0.118$ & $+0.644$ & 78.0 & $+0.083$ \\
\chem{HO_2} (R9) & 81.52 & $+0.008$ & $+0.852$ & 80.8 & $+0.084$ \\
\chem{HO_2} (R14) & 7.27 & $+0.371$ & $+0.635$ & 86.0 & $+0.115$ \\
\chem{HO_2} (sum) & 101.53 & $+0.062$ & $+0.805$ & 80.8 & $+0.087$ \\
\bottomhline
\end{tabular}
\belowtable{
\begin{list}{}{}
\item[$^\mathrm{a}$] continuum feature or component (\mbox{X-shooter}) or
  emission of given molecule and reaction in Tables~\ref{tab:chemFe} and
  \ref{tab:chemHO2} (WACCM)
\item[$^\mathrm{b}$] mean intensity from nighttime climatologies in
  Figs.~\ref{fig:featclim_I}, \ref{fig:simclim_Fe_I}, and
  \ref{fig:simclim_HO2_I}, or continuum component in Fig.~\ref{fig:contcomp}
\item[$^\mathrm{c}$] correlation coefficient for correlation with climatology
  of 595\,\unit{nm} feature in Fig.~\ref{fig:featclim_I}a
\item[$^\mathrm{d}$] correlation coefficient for correlation with climatology
  of 1,510\,\unit{nm} feature in Fig.~\ref{fig:featclim_I}b
\item[$^\mathrm{e}$] range of possible centroid emission heights from
  Q2DW-related analysis in Sect.~\ref{sec:heights} (\mbox{X-shooter}) and
  mean centroid emission heights from nighttime climatologies (WACCM)
\item[$^\mathrm{f}$] mean relative solar cycle effect for the intensity for a
  change of the solar radio flux averaged for 27 days by 100\,\unit{sfu} from
  the corresponding nighttime climatologies (plotted in
  Fig.~\ref{fig:featclim_SCE} for the \mbox{X-shooter}-related features)  
\end{list}
}
\label{tab:ressim}
\end{table*}

\begin{figure*}[t]
\includegraphics[width=17cm]{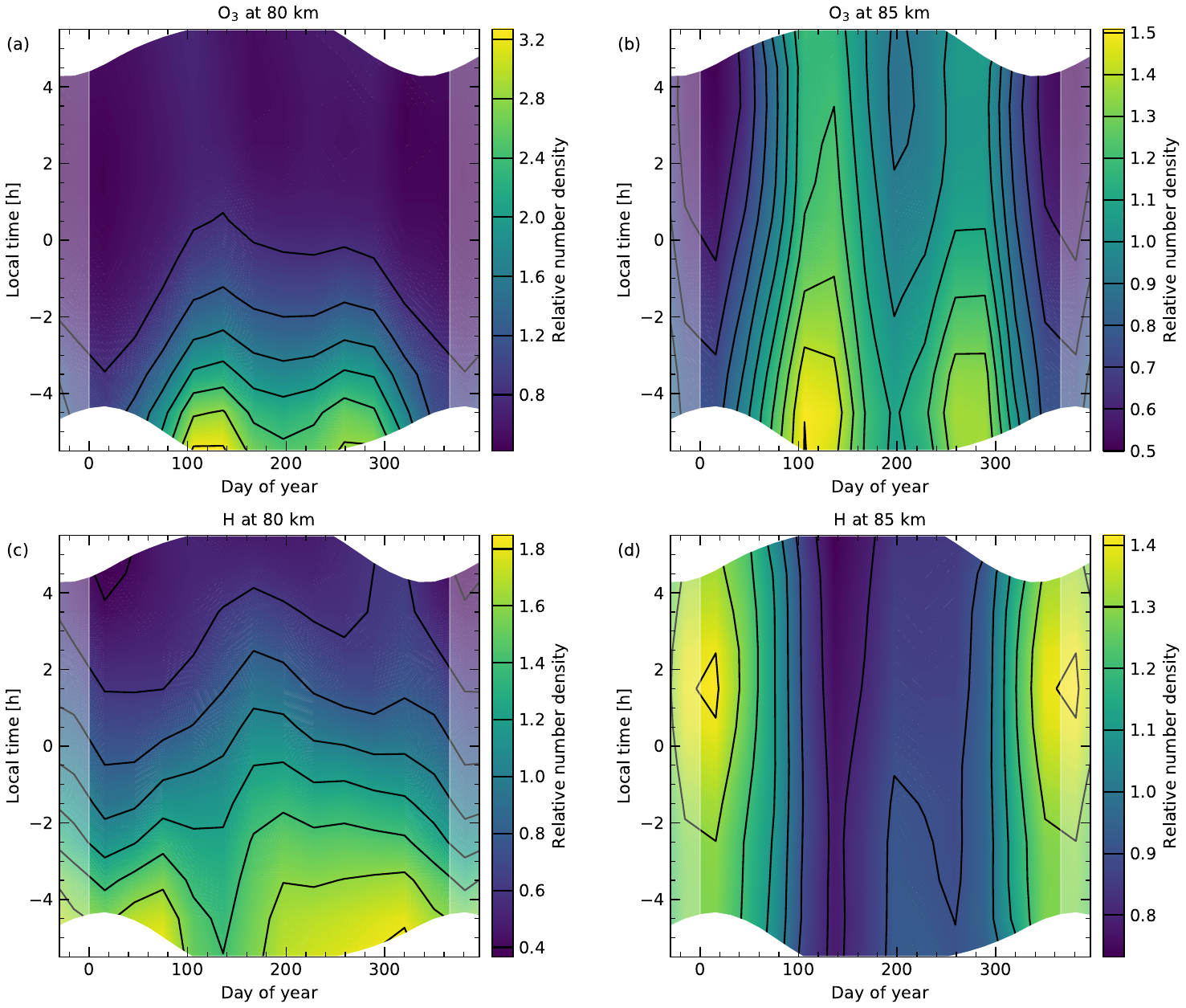}
\caption{Climatologies of WACCM-simulated relative number density at
  80\,\unit{km} (left) and 85\,\unit{km} (right) for the chemically important,
  strongly variable species \chem{O_3} (top) and \chem{H} (bottom) at
  24$^{\circ}$\,S and 70$^{\circ}$\,W. Sample and calculation of the
  climatologies were the same as for Fig.~\ref{fig:simclim_Fe_I}.}
\label{fig:simclim_O3_H}
\end{figure*}

Figure~\ref{fig:simprof} provides an overview of the typical nighttime
emission and density profiles of the different relevant species. Only profiles
close to local midnight were considered for the calculation of the mean
curves. The climatological variations with respect to local time and day of
year for relative intensity of \chem{FeO} and \chem{OFeOH} are shown in
Fig.~\ref{fig:simclim_Fe_I}. For the four \chem{HO_2}-related emission
processes listed in Table~\ref{tab:chemHO2}, the corresponding climatologies
are displayed in Fig.~\ref{fig:simclim_HO2_I}. The reference intensities
$\langle I \rangle$ for these climatologies are provided in
Table~\ref{tab:ressim}. They are compared to the corresponding results for
the \mbox{X-shooter}-based analysis, which involves the measurement of
individual continuum features and the derivation of continuum components. For
the intensity, the table also shows the correlation coefficients $r$ for the
correlation of the model climatologies of the different analysed emission
processes with the variability patterns of the measured continuum features at
595\,\unit{nm} (f06a) and 1,510\,\unit{nm} (f15a) that are displayed in
Fig.~\ref{fig:featclim_I}. The table also lists average centroid emission
heights $\langle h_\mathrm{cen} \rangle$ from the model-based nighttime
climatologies, compared with the ranges indicated from the Q2DW-related
analysis of the two continuum features in Sect.~\ref{sec:heights}. Finally,
the climatology-based effective solar cycle effect
$\langle \mathrm{SCE} \rangle$ is provided for the modelled and measured
intensities.

\subsubsection{\chem{FeO} and \chem{OFeOH} emission}\label{sec:FeO}

The best-known structure of the nightglow continuum is the peak at about
595\,\unit{nm}, which is identified to be caused by \chem{FeO} emission
\citep{evans10,saran11,gattinger11a,unterguggenberger17}. The related WACCM
climatology in Fig.~\ref{fig:simclim_Fe_I}a shows a primary maximum in May and
a secondary one in October, whereas the lowest nocturnal averages occur around
the turn of the year. The climatology for the 595\,\unit{nm} feature in
Fig.~\ref{fig:featclim_I}a shows a similar seasonal variability pattern with
only slight shifts in the peak positions. Although the moderate nocturnal
decrease between austral autumn and spring in the WACCM data is not present in
the measured climatology, the overall agreement is nevertheless satisfactory.
The correlation coefficient $r$ for the grid cells with significant nighttime
contribution is $+0.81$. Table~\ref{tab:ressim} also shows an average centroid
height for the \chem{FeO} emission of Reaction~\ref{eq:Fe+O3} of
87.9\,\unit{km} (see also Fig.~\ref{fig:simprof}a), which is located between
the low variability-based and maximum centroid emission heights from the
analysis of Q2DW-related variations of the 595\,\unit{nm} variations in
Sect.~\ref{sec:heights}. Even the minimum and maximum values of the simulated
climatology (which shows an increase with increasing LT) of 86.3 and
88.9\,\unit{km} are inside this interval. Moreover, the moderate positive
solar cycle effects from the modelled and measured climatologies of about
$+16$ and $+11$\% per 100\,\unit{sfu} agree quite well within their
uncertainties (Sects.~\ref{sec:modelsetup} and \ref{sec:SCE}, respectively).

Together with the spectral structure of the continuum in the VIS arm discussed
in Sects.~\ref{sec:meancont} and \ref{sec:contdecomp}, this appears to be a
robust identification of \chem{FeO} chemiluminescence. However, the
595\,\unit{nm} feature is only a very small part of the entire well-correlated
\chem{FeO}(VIS) component plotted in Fig.~\ref{fig:contcomp}.
Table~\ref{tab:ressim} lists a mean intensity of the measured feature of about
27\,\unit{R}, whereas the \chem{FeO}(VIS) component could emit about
2.9\,\unit{kR}. In contrast, the simulated mean intensity is only
170\,\unit{R} assuming a quantum yield (QY) of unity. This value is sufficient
for the main peak of the orange bands, which would imply a QY of about 16\%.
This percentage is consistent with the result of \citet{unterguggenberger17}
of $13 \pm 3$\,\%, which is also based on \mbox{X-shooter} measurements and
WACCM simulations, although the samples and analysis approaches differ. If
other features and more underlying continuum is added to the calculation of
the intensity, the simulated \chem{FeO} emission budget is consumed quite
fast. Adding just the continuum below the integration range of the
595\,\unit{nm} feature between 584 and 607\,nm leads to 148\,\unit{R} for the
mean continuum and 123\,\unit{R} if only the \chem{FeO}(VIS) component is
considered. Integration of the nightglow spectrum between 560 and
620\,\unit{nm} and subtraction of a constant flux that was measured at
500\,\unit{nm} returned 221\,\unit{R}. This kind of measurement was already
performed by \citet{saran11} for ESI spectra taken at Mauna Kea (see
Sect.~\ref{sec:intro}). They found intensities up to 157\,\unit{R} in two
nights. A simulation based on FeMOD \citep{gardner05} returned 61\,\unit{R},
which was then compared to a measured intensity of about 80\,\unit{R}.
Interestingly, their ratio of 1.3 for measurement vs. model is the same that
we obtain for our \mbox{X-shooter} and WACCM data. Moreover,
\citet{unterguggenberger17} stated that the \chem{FeO} main peak only
contributes about 3.9\% to the entire spectrum of the orange bands calculated
by \citet{gattinger11a}. For our case, this would be a total emission of
692\,\unit{R}. \citet{unterguggenberger17} also measured a fraction of
$3.3 \pm 0.8$\% of the main peak contribution to the emission between 500 and
720\,\unit{nm}. This is consistent with our results. We obtain 2.8\% for the
full nightglow continuum and 3.1\% for the \chem{FeO}(VIS) component.

Hence, the emissions of the apparent structures of the \chem{FeO} orange bands
in the nightglow continuum are already an issue for the model. Furthermore,
the simulated intensity is more than an order of magnitude too low if the
entire \chem{FeO}(VIS) component spectrum is compared. This result is
difficult to explain as WACCM agrees well with \chem{Fe} concentrations
measured by lidars \citep{feng13}. Moreover, the rate coefficient for
Reaction~\ref{eq:Fe+O3} has been measured in the laboratory \citep{helmer94}
and is close to the capture rate, i.e. the physical upper limit. A possible
alternative production pathway of \chem{FeO} would be the reaction of
relatively abundant \chem{FeOH} (see Fig.~\ref{fig:simprof}b) and \chem{H}.
However, this reaction is exothermic by only 61\,\unit{kJ\,mol^{-1}}, which is
not sufficient to produce the observed spectrum. In principle, the latter
would be possible by the sufficiently exothermic Reaction~R10 in
Table~\ref{tab:chemFe} that produces excited \chem{FeO_2}. However, our
simulation indicates that this emission is very faint
(Fig.~\ref{fig:simprof}a). The mean intensity in Table~\ref{tab:ressim} is
only 0.2\,\unit{R}.

The last imaginable \chem{Fe}-related emission process would be related to
Reaction~\ref{eq:FeOH+O3} of \chem{FeOH} and \chem{O_3}. The resulting
\chem{OFeOH} radiation could cover the entire wavelength regime of the
\chem{FeO}(VIS) component, although wavelengths above 500\,\unit{nm} would be
more likely (Sect.~\ref{sec:modelsetup}). As listed by Table~\ref{tab:ressim},
this nightglow process could produce up to 220\,\unit{R}, which is more than
in the case of \chem{FeO}. In order to be relevant for the \chem{FeO}(VIS)
component, the variability needs to be similar to the climatology of the
595\,\unit{nm} feature in Fig.~\ref{fig:featclim_I}a. Indeed, the pattern in
Fig.~\ref{fig:simclim_Fe_I}b is very similar. The correlation coefficient is
$+0.87$, which is even higher that in the case of the simulated \chem{FeO}
emission. Compared to the latter, the \chem{OFeOH} climatology shows a more
prominent secondary peak in austral spring and a later nocturnal maximum.
However, the discrepancies appear to be small enough that both emissions
could contribute to the same NMF component. A clear difference are the
emission profiles as shown in Fig.~\ref{fig:simprof}a. The mean nighttime
centroid emission height for \chem{OFeOH} is 82.3\,\unit{km}, which is
5.6\,\unit{km} lower than in the case of \chem{FeO}. It is also clearly lower
than the derived range for the 595\,\unit{nm} feature. This is not an issue if
this feature is mostly produced by the \chem{FeO}-related
Reaction~\ref{eq:Fe+O3} and the contribution of Reaction~\ref{eq:FeOH+O3}
between 564 and 680\,\unit{nm} is rather small in general. The latter
constraint results from the OSIRIS-based emission profile with a peak at about
87\,\unit{km} (Sect.~\ref{sec:heights}) that was derived for this wavelength
range by \citet{evans10}. The effective solar cycle effect as shown in
Table~\ref{tab:ressim} should also mainly be determined by
Reaction~\ref{eq:Fe+O3}. As the value for Reaction~\ref{eq:FeOH+O3} is also
not far away from the measurement for the 595\,\unit{nm} feature, the
contribution of both emission processes cannot be distinguished as in the case
of the centroid heights.

These results suggest that \chem{OFeOH} emission could significantly
contribute to the \chem{FeO}(VIS) component, although the emission spectrum
(if any) is unknown, and the rate coefficient has not been measured. Here, it
is set to the collision frequency, and a QY of unity is assumed to provide an
upper limit to the contribution of this reaction. Despite this, the summed
emission of Reactions~\ref{eq:Fe+O3} and \ref{eq:FeOH+O3} is still too low to
explain the full nightglow continuum in the \mbox{X-shooter} VIS arm. So far,
we have not succeeded to identify another metal-related or any reaction that
could explain the missing emission. In any case, it would be essential that
the climatological variability pattern is mainly determined by the variations
of \chem{O_3}, which is a reactant in both \chem{Fe}-related reactions. As
shown by Fig.~\ref{fig:simclim_O3_H}b for an altitude of 85\,\unit{km}, the
semiannual pattern with maxima near the equinoxes and the main minimum in
austral summer is a clear indicator of \chem{O_3} density changes in the
mesopause region above Cerro Paranal.

\subsubsection{\chem{HO_2} emission}\label{sec:HO2}

The climatologies of the \chem{Fe}-related emissions in
Fig.~\ref{fig:simclim_Fe_I} are very different from the variations of the
1,510\,\unit{nm} feature in Fig.~\ref{fig:featclim_I}b. The correlation
coefficients are close to 0 (Table~\ref{tab:ressim}). Hence, we can focus our
discussion of the possible emitter of the peak at 1,510\,\unit{nm} and the
other structures of the \chem{X}(NIR) continuum on \chem{HO_2}, which appears
to be the primary candidate for these emission features according to the
discussion in Sect.~\ref{sec:contdecomp}.

The mean intensities in Table~\ref{tab:ressim} show that there should be
sufficient emission to produce the entire \chem{X}(NIR) continuum. For the sum
of the three reactions of about 102\,\unit{kR}, an effective QY of
only about 12\% would be necessary if we neglect possible emission beyond
1,800\,\unit{nm} (Sect.~\ref{sec:intclim}). If we only consider
Reaction~\ref{eq:H+O2+M}, it would be 14\%. Although the individual QYs could
be quite different, the reaction involving \chem{H} and \chem{O_2} appears to
be the dominating pathway. The intensity discrepancy is fairly large (see also
Fig.~\ref{fig:simprof}c).

The climatology of Reaction~\ref{eq:H+O2+M} also indicates the best
correlation with the climatology of the 1,510\,\unit{nm} feature in
Fig.~\ref{fig:featclim_I}b. The corresponding $r$ value is about $+0.85$,
whereas these coefficients are close to $+0.64$ for the two other reactions.
The reasons for these differences are illustrated in
Fig.~\ref{fig:simclim_HO2_I}. All reactions show a decrease of the intensity
in the course of the night. However, the relatively constant rate of this
decrease for (b) and (c) provides a much better agreement with the climatology
of the 1,510\,\unit{nm} feature than the steep exponential drop of the
intensity from Reaction~\ref{eq:HO2+O2a} at the beginning of the night in (a).
The latter is related to the decay of the
\chem{O_2}($\mathrm{a}^1\Delta_{\mathrm{g}}$) population produced by \chem{O_3}
photolysis at daytime. At later LTs,
\chem{O_2}($\mathrm{a}^1\Delta_{\mathrm{g}}$) is then mainly produced by
Reaction~R15 in our model.

\begin{figure}[t]
\includegraphics[width=8.3cm]{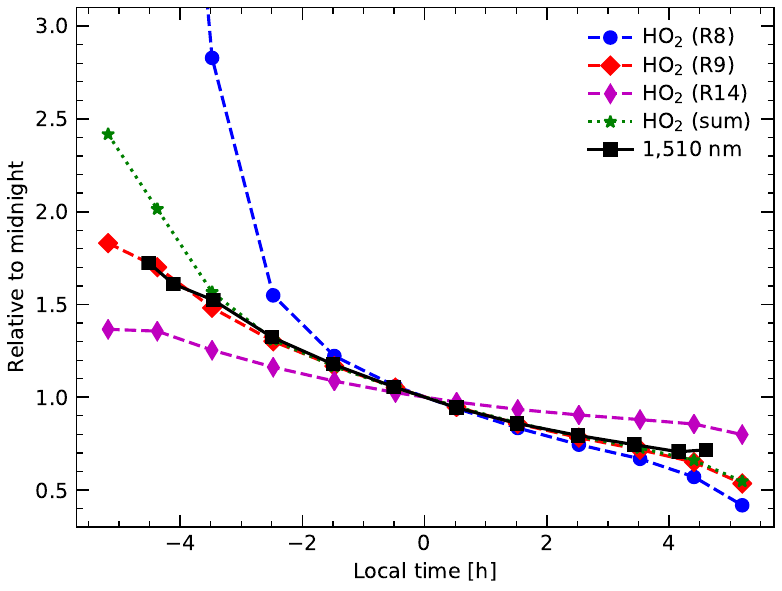}
\caption{Climatology-based mean nighttime trend of WACCM \chem{HO_2}
  intensities for the individual reactions in Table~\ref{tab:chemHO2} and the
  sum of them in comparison to the result for the climatology of the
  1,510\,\unit{nm} emission plotted in Fig.~\ref{fig:featclim_I}b. All curves
  are provided relative to the mean intensity of the two data points close to
  midnight. The highest value for the curve for Reaction~\ref{eq:HO2+O2a} (see
  legend), which is partly outside the plotted range, is 10.0. The given local
  times are the averages of the data sets that were used to calculate the
  climatological grid. The shorter time coverage of the curve for the
  1,510\,\unit{nm} feature reflects the lack of observations with central LTs
  close to twilight.}
\label{fig:noctvar_HO2}
\end{figure}

For an easier comparison, Fig.~\ref{fig:noctvar_HO2} shows the mean nocturnal
trends from the different climatologies scaled to midnight.
Reaction~\ref{eq:HO2+O2a} is a clear outlier. Moreover, the intensity from
Reaction~R14 appears to decrease too slowly compared to the curve for the
1,510\,\unit{nm} emission. On the other hand, Reaction~\ref{eq:H+O2+M}
seems to match almost perfectly. Of course, this could also be coincidence to
some extent. A check of the monthly nocturnal trends also indicates clear
deviations. The trend for the simulated emission tends to be steeper from
November to March and flatter from April to October. Although possible
systematic deviations of the model are difficult to estimate, this result
favours Reaction~\ref{eq:H+O2+M}. The large contribution of the latter to the
summed intensity means that the nocturnal trends of the summed intensity and
due just to Reaction~\ref{eq:H+O2+M} look very similar in
Fig.~\ref{fig:noctvar_HO2}. The only noteworthy discrepancy occurs at the
beginning of the night due to the high intensity related to
Reaction~\ref{eq:HO2+O2a} (see also Fig.~\ref{fig:simclim_HO2_I}d). However,
the difference is still relatively small at the earliest data point for the
1,510\,\unit{nm} emission. Hence, some contribution of this reaction to the
total emission cannot be excluded, but it should not be significantly higher
than modelled for equal QYs. This statement refers to the branch of the
\chem{HO_2} production by Reaction~\ref{eq:HO2+O2a} related to daytime
\chem{O_3} photolysis. The mean intensity after midnight is only
5.0\,\unit{kR}, i.e. 39\% of the mean of the full nighttime climatology.
However, this value is quite uncertain as our discussion of the nocturnal
\chem{O_2}($\mathrm{a}^1\Delta_{\mathrm{g}}$) generation in
Sect.~\ref{sec:modelsetup} illustrates.

In order to obtain a rough estimate of the quality of our assumptions, we
made a simple conversion of the \chem{O_2}($\mathrm{a}^1\Delta_{\mathrm{g}}$)
densities from WACCM (mean midnight profile in Fig.~\ref{fig:simprof}d) to
emission rates assuming a QY of unity and using an effective Einstein-A
coefficient of $2.28 \times 10^{-4}$\,\unit{s^{-1}} for the
\mbox{\chem{O_2}(a-X)(0-0)} band emission at 1,270\,\unit{nm} \citep{noll16}.
We then compared the vertically integrated emission rates with those from
SABER measurements in the corresponding channel. For this purpose, we used
the 4,496 profiles that were collected for Cerro Paranal by \citet{noll17}.
Next, we performed a similar analysis of the nocturnal intensity trend as
described by \citet{noll16}, i.e. we fitted an exponential function and a
constant for the time after sunset. As the WACCM data set is quite large,
we performed this for each month separately and averaged the results. The
fit functions were scaled to the mean of the second half of the night. For
the intensities for the entire column above 40\,\unit{km}, these reference
constants are 29.7\,\unit{kR} for WACCM and 102.0\,\unit{kR} for SABER.
However, the vertical emission distribution for \mbox{\chem{O_2}(a-X)(0-0)}
with a centroid emission height of about 89\,\unit{km} around midnight
\citep{noll16} is quite different from the WACCM-based profile shown in
Fig.~\ref{fig:simprof}d. Hence, we restricted the comparison to the height
range between 80 and 85\,\unit{km}. Then, we obtain reference intensities of
18.4\,\unit{kR} (62\%) for WACCM and 15.4\,\unit{kR} (15\%) for SABER. For
the exponential component, we fitted time constants of about 74 and
76\,\unit{min}, which are very close to the radiative lifetime of about
73\,\unit{min} from the Einstein-A coefficient. For the entire vertical
column, the lifetimes decrease to 55 and 50\,min in agreement with the
results from \citet{noll16}. The change is related to the higher impact of
collisional deactivation at lower altitudes. In any case, the good agreement
of the WACCM and SABER time constants indicate that the related temporal
variations are well simulated by WACCM. For the integration from 80 to
85\,\unit{km}, the intensities of the exponential component are
71.8\,\unit{kR} for WACCM and 169.9\,\unit{kR} for SABER 60\,\unit{min} after
sunset. As these values are 21 to 22\% of the corresponding intensities for
the whole column, WACCM also performs quite well with respect to the vertical
distribution of \mbox{\chem{O_2}(a-X)(0-0)} produced by \chem{O_3} photolysis.
However, the WACCM-related intensity is clearly lower. On the other hand,
the corresponding intensity for the nighttime production is somewhat higher
than the value from SABER. If we assume a constant ratio of the WACCM and
SABER intensities for the entire night, then the branching ratio in
Reaction~R15 needs to be lowered to about 33\%. Of course, the uncertainties
of this comparison are quite high. For example, the height-dependent impact of
collisions can play a role. However, the results appear to show that a
distinctly higher contribution of Reaction~\ref{eq:HO2+O2a} to the generation
of excited \chem{HO_2} in the night than simulated is unlikely, which
strengthens the importance of Reaction~\ref{eq:H+O2+M}.

The climatology of the 1,510\,\unit{nm} feature in Fig.~\ref{fig:featclim_I}b
indicates semiannual variations with a main maximum in austral summer and a
secondary maximum in winter. The \chem{HO_2}-related climatologies in
Fig.~\ref{fig:simclim_HO2_I} show discrepancies with respect to the seasonal
variations. Nevertheless, the highest intensities are also present in summer
for Reaction~\ref{eq:H+O2+M}. If the first few hours of the night are
excluded, this also appears to be true for Reaction~\ref{eq:HO2+O2a}. The
climatologies of these reactions are also characterised by minimum intensities
in May to June slightly depending on LT. The seasonal pattern of Reaction~R14
is very different as it indicates a semiannual oscillation with maxima around
the equinoxes, which explains the low correlation coefficient in
Table~\ref{tab:ressim}. This pattern is obviously related to the important
role of \chem{O_3} in the production of \chem{HO_2}. The similarities are
clearly visible in the top row of Fig.~\ref{fig:simclim_O3_H}, where the
climatologies of the \chem{O_3} densities at 80 and 85\,\unit{km} are
displayed. Moreover, the bottom row of the same figure, which shows \chem{H}
for the same heights, explains the location of the seasonal maxima and minima
for the other reactions. This makes sense as \chem{H} is the only strongly
variable reactant in the production of \chem{HO_2} via
Reaction~\ref{eq:H+O2+M}. As Reaction~\ref{eq:HO2+O2a} requires the previous
generation of \chem{HO_2}, there is also an impact of the \chem{H} variability
on this reaction. It also explains why the climatology of Reaction~R14 still
better correlates with the variations of the 1,510\,\unit{nm} feature than the
595\,\unit{nm} feature (Table~\ref{tab:ressim}). Hence, this pathway (despite
some similarities with the \chem{Fe}-related emissions discussed in
Sect.~\ref{sec:FeO}) does not appear to be a suitable contributor to the
\chem{FeO}(VIS) continuum component.

\begin{figure}[t]
\includegraphics[width=8.3cm]{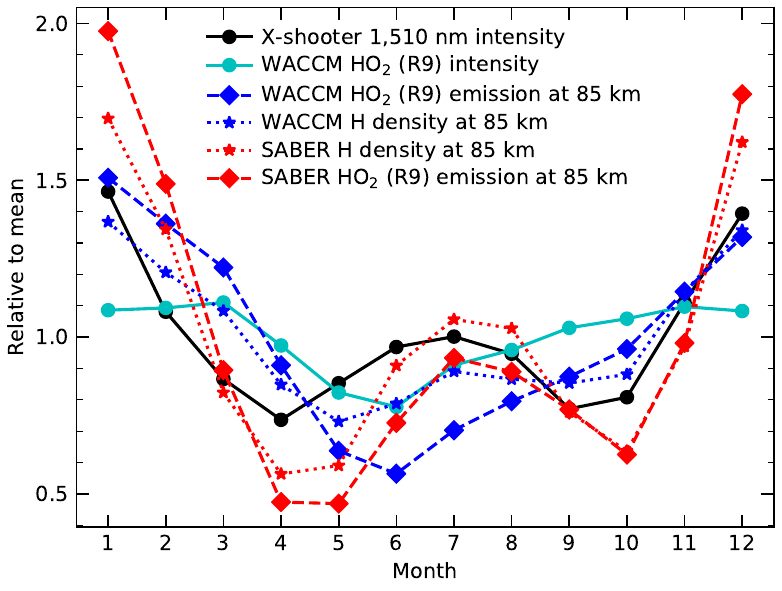}
\caption{Seasonal variations of monthly averages of time series of the
  \mbox{X-shooter}-based 1,510\,nm intensity (7,931 30\,\unit{min} bins), the
  WACCM \chem{HO_2} intensity for the dominating Reaction~\ref{eq:H+O2+M}
  (92,064 time steps), the corresponding \chem{HO_2} emission rate and
  \chem{H} density at 85\,\unit{km}, and the SABER-based \chem{HO_2} emission
  rate and \chem{H} density at the same altitude (16,079 profiles) for Cerro
  Paranal. The different curves (see legend) are provided relative to the mean
  for the 12 months.}
\label{fig:seasvar_HO2_H}
\end{figure}

In order to also identify \chem{H} as the main driver of the climatological
variations of the 1,510\,\unit{nm} feature, we need to find the reason for
the missing secondary maximum in the WACCM simulation in winter.
\citet{mlynczak14} performed global \chem{H} retrievals based on the
2.1\,\unit{\mu{}m} channel of SABER \citep{russell99}. As a result, they found
semiannual seasonal variations at 84\,\unit{km} and a latitude of
16.5$^{\circ}$\,S with a secondary maximum in austral summer. This pattern was
confirmed by a comparison with older results at 20$^{\circ}$\,S from
\citet{thomas90} based on measurements with the near-IR spectrometer on the
Solar Mesosphere Explorer. \citet{mlynczak18} improved the \chem{OH}-based
retrieval algorithm for \chem{O} and \chem{H}, which reduced the densities.
We already used this data set for the years 2002 to 2014 for an analysis of
\chem{K} emissions above Cerro Paranal \citep{noll19}. This allows us to
reuse these data for the study of the seasonal \chem{H} variations. For this
purpose, we calculated simple monthly mean values for the SABER and WACCM time
series. For an altitude of 85\,\unit{km}, Fig.~\ref{fig:seasvar_HO2_H}
confirms that there is a weak secondary maximum in the SABER \chem{H} density.
Moreover, the primary peak in summer is more pronounced than in the WACCM
simulation. In the next step, we compared \chem{HO_2} emission rates. For the
dominating Reaction~\ref{eq:H+O2+M}, these were calculated from the SABER
densities of \chem{H} and air and the rate coefficients in
Table~\ref{tab:chemHO2}. Only the weakly-varying \chem{O_2} volume mixing
ratio was taken from WACCM. The results for 85\,\unit{km} are also presented
in Fig.~\ref{fig:seasvar_HO2_H}. The deviations from the variations of
\chem{H} are small and similar for both data sets, which illustrates the
dominating role of \chem{H} for the \chem{HO_2} emission variability. The
comparison of the WACCM \chem{HO_2} emission at 85\,\unit{km} and for the
entire vertical column indicates a smoothing of the seasonal pattern in the
latter case. We cannot test the same for the SABER data as the noise in the
\chem{H} retrievals quickly increases at lower altitudes due to the weaker
emission of the \mbox{\chem{OH}(8-6)} and \mbox{\chem{OH}(9-7)} bands that
were essentially used for the derivation of the \chem{H} density
\citep{mlynczak18}. This is also the main reason why the decreasing nocturnal
trend in the \chem{HO_2} emission could not be studied in this way.
Figure~\ref{fig:simclim_O3_H} indicates that this trend is obviously generated
distinctly lower than 85\,\unit{km}. Nevertheless, if we assume that the
smoothing in the seasonal pattern in the WACCM data is similar for the
observations, it is likely that we would obtain a variability structure that
resembles the curve for the 1,510\,\unit{nm} feature plotted in
Fig.~\ref{fig:seasvar_HO2_H}. Hence, the differences between modelled and
observed emission variations appear to be mostly caused by the WACCM-based
reproduction of temporal changes in the \chem{H} density.

Table~\ref{tab:ressim} shows the mean centroid emission altitudes from the
corresponding climatologies of the different production processes of
\chem{HO_2} emission. Only the height of about 81\,\unit{km} for
Reaction~\ref{eq:H+O2+M} is in the range from about 80 to 84\,\unit{km} for
the average profile derived from the \mbox{X-shooter}-based analysis in
Sect.~\ref{sec:heights}. The agreement looks even more promising if it is
considered that \chem{O}, which is crucial for the upper limit of
84\,\unit{km}, is not involved in the chemical production of excited
\chem{HO_2} for this pathway. The simulated climatological variations show
increasing heights during the night and the highest values in austral summer
with maximum deviations from the mean lower than 2\,\unit{km} in most cases.
The emission of Reaction~\ref{eq:HO2+O2a} just reaches 80\,km close to
sunrise, whereas the centroid heights are even below 70\,\unit{km} at the
beginning of the night due to the high concentrations of
\chem{O_2}($\mathrm{a}^1\Delta_{\mathrm{g}}$) related to daytime \chem{O_3}
photolysis in the lower mesosphere \citep{noll16}. Reaction~R14 shows a
similar variability pattern as Reaction~\ref{eq:H+O2+M} but, with a mean value
of 86\,\unit{km}, the emission is too high. The reason is the impact of
\chem{O_3}, which shows its density peak at a higher altitude than \chem{H} in
the mesopause region in Fig.~\ref{fig:simprof}d. This plot also indicates that
the density distribution of all \chem{HO_2} peaks at 78\,\unit{km}, which is
lower than the emission maximum of excited \chem{HO_2} shown in
Fig.~\ref{fig:simprof}c.

Finally, Table~\ref{tab:ressim} shows that at least the results for
Reactions~\ref{eq:HO2+O2a} and \ref{eq:H+O2+M} appear to agree with a positive
but weak solar cycle effect that was derived for the 1,510\,\unit{nm} feature
in Sect.~\ref{sec:SCE}. Although the uncertainties are of the order of several
per cent, the almost doubled deviation from the measured value compared to the
other reactions, make Reaction~R14 less likely with respect to the response to
solar activity.

Our investigation of the three proposed production mechanisms for excited
\chem{HO_2} shows promising results. It is likely that \chem{HO_2} is the
radiating molecule \chem{X} that produces the 1,510\,\unit{nm} feature and the
strongly correlated \chem{X}(NIR) continuum. Moreover, the recombination
reaction of \chem{H} and \chem{O_2} with participation of an additional
collision partner is the most probable production process for excited
\chem{HO_2}. All investigated properties such as total emission, nocturnal and
seasonal variations, emission heights, and solar cycle effect point to this
interpretation. It also helps that chemiluminescence by this mechanism was
already observed in the laboratory between about 800 and 1,550\,\unit{nm}
\citep{holstein83}, although an extension to higher wavelengths would be
important to test the presence of the 1,620\,\unit{nm} feature
(Fig.~\ref{fig:contcomp}). Reaction~\ref{eq:HO2+O2a} that involves
\chem{O_2}($\mathrm{a}^1\Delta_{\mathrm{g}}$) can only be a minor emission
source, otherwise the observed nocturnal trend would not fit. It is
also more challenging to produce emission below 1,270\,\unit{nm} for this
process. Finally, Reaction~R14 that involves \chem{O_3} shows clear
discrepancies in the emission properties which can only be tolerated if the
contribution to the total emission is very small.

\conclusions[Conclusions]  
\label{sec:conclusions}

Our analysis of the nightglow (pseudo-)continuum with high-quality
\mbox{X-shooter} data essentially reveals two contributions in the wavelength
range between 300 and 1,800\,\unit{nm} if remnants from different \chem{O_2}
bands are excluded.

Our results of the correlation analysis of continuum structures and
non-negative matrix factorisation (NMF) of the continuum variability show that
the peak at 595\,\unit{nm} is well correlated with other features and the
underlying continuum in a wide wavelength range, especially between about 500
and 900\,\unit{nm}. The variations as mainly studied for the feature at
595\,\unit{nm} reveal a climatology with a mixture of semiannual and annual
oscillation with a main maximum in April/May and a main minimum in January
that confirms previous results based on a smaller sample. For the first time,
we estimated the effective solar cycle effect and found a weak positive
correlation. Using an approach for the estimate of effective emission heights
based on the analysis of a strong passing Q2DW that was initially developed
for \chem{OH} lines, we obtained a range for the mean centroid height between
about 85 and 89\,\unit{km}.

In previous studies, the feature at 595\,\unit{nm} was identified as the main
peak of the \chem{FeO} orange bands. Our simulations of the chemiluminescence
from the reaction of \chem{Fe} and \chem{O_3} with WACCM can reproduce most of
the measured properties of the emission, which suggests that the NMF component
dominating the \mbox{X-shooter} VIS arm could have contributions from various
\chem{FeO} bands. However, WACCM returns a maximum mean intensity of only
170\,\unit{R}, whereas the whole correlated spectrum could have
2.9\,\unit{kR}. We discovered that potential \chem{OFeOH} emission (with
unknown spectral distribution) from the reaction between \chem{FeOH} and
\chem{O_3} would have a very similar climatology according to our simulations.
Nevertheless, this reaction would only add up to 220\,\unit{R}. Therefore, a
major discrepancy remains. If there is another emitter, the basic precondition
for a good correlation with the 595\,\unit{nm} feature appears to be that the
variability is mainly determined by \chem{O_3}. 

The second continuum component dominates the \mbox{X-shooter} NIR arm. In
particular, a strong narrow peak at about 1,510\,\unit{nm} and a secondary
feature at about 1,620\,\unit{nm} were found, which indicates a complex band
system. Our best estimate of the average intensity of the entire system is
about 12\,\unit{kR}. The seasonal variations with maxima near the solstices
are actually in opposition to those of the 595\,\unit{nm} feature. There is
also a clear decrease of the intensity in the course of the night for the
entire year. The solar cycle effect is only weakly positive and the average
effective emission height appears to be most likely between about 80 and
84\,\unit{km} for the 1,510\,\unit{nm} feature.

The most promising candidate for the emitter is \chem{HO_2}. Existing near-IR
spectra from the laboratory suggest that the 1,510\,\unit{nm} feature could be
the vibrational \mbox{(200-000)} transition of the electronic ground state
$^2\mathrm{A}^{\prime\prime}$. There would also be an explanation of the
enhanced emission near 1,270\,\unit{nm}, where only a part appears to be due
to residual \chem{O_2} emission. Other features from the experiments could not
be checked due to gaps in our continuum spectrum. We investigated different
potential production processes of excited \chem{HO_2} with WACCM. The
recombination reaction between \chem{H} and \chem{O_2} under participation of
another collision partner showed the best performance. It is the main
production process of \chem{HO_2} in the mesosphere. With a modelled maximum
mean radiance of 82\,\unit{kR}, a moderate quantum yield of the reaction would
be sufficient to produce the continuum in the \mbox{X-shooter} NIR arm.
Moreover, this process indicates the best agreement with respect to the
climatological variations. Remaining discrepancies (especially a missing
secondary peak in austral winter) can be explained by deviations of the
modelled \chem{H} densities from those of SABER-based retrievals. The
simulated weak solar cycle effect and the average centroid emission height of
about 81\,\unit{km} also show good agreement. Finally, the observed
chemiluminescence in the laboratory for this mechanism indicated emission down
to about 800\,\unit{nm}, which is consistent with the shape of the derived
continuum component. As wavelengths above about 1,550\,\unit{nm} were not
studied in the only known laboratory experiment, there is no evidence of the
existence of the 1,620\,\unit{nm} feature, so far. The other studied potential
emission processes appear to be much less efficient. Relatively weak or even
negligible emission rates are probably related to the direct radiative
recombination of \chem{H} and \chem{O_2}, the reaction of \chem{H} and
\chem{O_3}, and collisions of \chem{HO_2} with
\chem{O_2}(a$^1\Delta_{\mathrm{g}}$). The latter would produce a steep decline
of the emission after dusk due to the decay of the population of excited
\chem{O_2} molecules produced by \chem{O_3} photolysis, which is not observed
in the \mbox{X-shooter} data. The reaction involving \chem{H} and \chem{O_3}
would generate a very different seasonal variability as observed.

The intriguing discoveries of this study will certainly stimulate further
investigations for a better understanding of the chemistry and dynamics in the
Earth’s mesopause region. The origin of the whole VIS-arm continuum still
needs to be solved. The study also revealed that the nighttime production of
\chem{O_2}(a$^1\Delta_{\mathrm{g}}$) is not understood well, although these
excited molecules are essential for the strong emission at 1,270\,\unit{nm}.
These examples illustrate that there are still many things at these altitudes
that we do not know.



\dataavailability{The basic \mbox{X-shooter} data for this project originate
  from the ESO Science Archive Facility at \mbox{http://archive.eso.org}
  and are related to various observing programmes that were carried out
  between October 2009 and September 2019. The raw spectra were processed 
  (using the corresponding calibration data) and then analysed. The NIR-arm
  data were already used for the study of \chem{OH} emission lines
  \citep{noll22,noll23}. Some results of these investigations, which are
  available via the public repository Zenodo \citep{noll22ds,noll23ds1}, were
  also considered for this study. We performed dedicated WACCM6 runs with
  modified chemistry for the years from 2003 to 2014. The crucial results are
  stored at the University of Leeds. We also made use of TIMED/SABER data sets
  that were already collected for previous studies for Cerro Paranal from the
  SABER website at \mbox{http://saber.gats-inc.com}. These are the v2.0
  products from 2002 to 2015 analysed by \citet{noll17} and the improved
  \chem{O} and \chem{H} retrievals described by \citet{mlynczak18} for the
  years 2002 to 2014 that were used by \citet{noll19}. Results from the study
  of v2.0 products from January and February 2017 by \citet{noll22} were also
  considered. A comprehensive collection of data of our analysis (especially
  with respect to the plotted data) is provided by Zenodo at
  \mbox{https://zenodo.org/record/8335836} \citep{noll23ds2}.}













\authorcontribution{SN designed and organised the project, performed the
  preparation and analysis of the \mbox{X-shooter} spectra, visualised the
  results based on \mbox{X-shooter}, WACCM, and SABER data, and is the main
  author of the paper text. The co-authors contributed to the improvement of
  the paper content. Moreover, JP designed the WACCM runs, investigated the
  involved chemistry, and significantly influenced the scientific discussion.
  WF performed the WACCM simulations and a preliminary analysis of the data.
  KK significantly contributed to the discussion of the chemistry. In
  particular, he first proposed \chem{HO_2} as possible emitter. WK carried
  out the basic processing of the \mbox{X-shooter} spectra. CS contributed to
  the discussion of the \mbox{X-shooter}-based analysis. MB was involved in
  the management of the project. SK managed the infrastructure for the
  processing and storage of the \mbox{X-shooter} data and contributed to the
  discussion of the measured continuum features.}

\competinginterests{Co-author John Plane is a member of the editorial board of
  Atmospheric Chemistry and Physics. The authors do not have other competing
  interests to declare.} 


\begin{acknowledgements}
  Stefan Noll was financed by the project \mbox{NO\,1328/1-3} of the German
  Research Foundation (DFG). John Plane and Wuhu Feng were supported by grant
  NE/T006749/1 from the UK Natural Environment Research Council. Konstantinos
  Kalogerakis acknowledges financial support from the US National Aeronautics
  and Space Administration (NASA grant 80NSSC23K0694) and the National Science
  Foundation (NSF grant AGS-2009960). We thank Sabine M\"ohler from ESO for
  her support with respect to the \mbox{X-shooter} calibration data. 
\end{acknowledgements}







\bibliographystyle{copernicus}
\bibliography{Nolletal2023b}

\end{document}